\documentclass[%
reprint,
amsmath,amssymb,
aps,
pre,
]{revtex4-2}
\usepackage{xcolor}
\usepackage{commath}
\usepackage{multirow}
\usepackage{graphicx}
\usepackage{subfigure}
\usepackage{dcolumn}
\usepackage{bm}
\usepackage{hyperref}
\usepackage[english]{babel}
\newcommand{\ee}{\mathrm{e}}
\newcommand{\ER}{Erd\H{o}s-R\'enyi}
\graphicspath{{figs/}}

\newcommand{\cut}[1]{{}}
\begin{document}

	\preprint{APS/123-QED}
	
	\title{Scalable Node-Disjoint and Edge-Disjoint Multi-wavelength Routing}
	
	\author{Yi-Zhi Xu$^1$,  Ho Fai Po$^2$, Chi Ho Yeung$^{2}$ and David Saad$^1$}
	\affiliation{$^1$The Nonlinearity and Complexity Research Group, Aston University, Birmingham B4 7ET, United Kingdom \\$^2$Department of Science and Environmental Studies, The Education University of Hong Kong, 10 Lo Ping Road, Taipo, Hong Kong. }
	
	\date{\today}

	\begin{abstract}
		Probabilistic message-passing algorithms are developed for routing transmissions in multi-wavelength optical communication networks, under node and edge-disjoint routing constraints and for various objective functions. Global routing optimization is a hard computational task on its own but is made much more difficult under the node/edge-disjoint constraints and in the presence of multiple wavelengths, a problem which dominates routing efficiency in real optical communication networks that carry most of the world's Internet traffic. The scalable principled method we have developed is exact on trees but provides good approximate solutions on locally tree-like graphs.  It accommodates a variety of objective functions that correspond to low latency, load balancing and consolidation of routes, and can be easily extended to include heterogeneous signal-to-noise values on edges and a restriction on the available wavelengths per edge. It can be used for routing and managing transmissions on existing topologies as well as for designing and modifying optical communication networks. Additionally, it provides the tool for settling an open and much debated question on the merit of wavelength-switching nodes and the added capabilities they provide. The methods have been tested on generated networks such as random-regular, Erd\H{o}s R\'{e}nyi and power-law graphs, as well as on the UK and US optical communication networks. They show excellent performance with respect to existing methodology on small networks and have been scaled up to network sizes that are beyond the reach of most existing algorithms. 
	\end{abstract}
	
	\maketitle

	\maketitle
	

	\section{Introduction}
	Optical communication networks underpin the global digital communications infrastructure and carry most of the Internet traffic. They comprise thousands of kilometers of optical fibers, organized in a complex web of constituent sub-networks including the Internet backbone, Metro access and Data Center farms. The exponential growth in Internet traffic and energy consumption threatens to overload the existing infrastructure and a capacity crunch is looming~\cite{Ellis2016}. Not only that a matching growth in infrastructure is infeasible, it raises fundamental questions on the ultimate capacity of optical communication networks and the manner in which we could optimize their use. The next-generation digital infrastructure has to offer flexibility, low latency, high network throughput and resilience. 
	
	One of the key requirements is the routing and wavelength assignment (RWA) for all traffic demands across this complex heterogeneous network in a way that optimizes a given objective function, be it low latency, high throughput or resilience~\cite{RWAreview}. Each optical fiber carries information using light of one or many wavelengths. The latter uses, among others, dense wavelength-division multiplexing (DWDM) methods that employ as many as $80$--$160$ channels of different laser wavelengths~\cite{Corcoran2020}. The main constraint in the RWA is that any complete individual route, from source to destination, uses the same single wavelength and that two separate routes using the same wavelength cannot shares the same fiber. This constraint makes the corresponding mathematical problem hard to solve in general.
	
	Route optimization in optical communication networks can be mapped onto the hard computational problem of edge-disjoint  routing on a graph, where transceivers (transmitter-receiver) are mapped to vertices (or nodes) and fibers to edges (or links). Given that routes are constrained to be contiguous and interaction between paths is non-localized, local optimization methods are insufficient and global optimization is required. Globally optimal routing of multiple messages or vehicles given a general objective function is a computationally-hard constraint satisfaction problems on its own and has been addressed in the physics literature using scalable and distributed message passing approximation techniques, inspired by statistical physics methodology~\cite{yeung2012competition,yeung2013physics,yeung2019coordinating,Po2021}. Moreover, similar techniques have been suggested also for addressing the \emph{single-wavelength} Node-Disjoint Paths (NDP)~\cite{de2014shortest} and Edge-Disjoint Paths (EDP)~\cite{altarelli2015edge} problems where multiple paths of different origin-destination pairs cannot share nodes or edges on a graph, respectively. Generally, both optimization tasks are within the class of NP-hard combinatorial problems~\cite{karp1972,garey1979computers,Korte2012,Erlebach2006} and the approximation offered by message passing techniques work well. However, the existing methods developed for single-wavelength routing become intractable in the presence of multiple wavelengths, making them inapplicable for realistic scenarios.
	
	It is worthwhile noting that in some extreme cases these hard computational problems become polynomial in the single wavelength case as discussed in~\cite{altarelli2015edge}. For instance, when the number of origin-destination pairs is low with respect to the systems size~\cite{ROBERTSON199565} and where all origin-destination pairs are identical~\cite{Vygen1994,VYGEN199583}. Nevertheless, the general problems of NDP and EDP routing on graphs are computationally hard even in the single wavelength case and a variety of methods have been used to address them in the context of optical communication networks and more general problems. Alongside established methods, such as  integer/linear programming and its variants~\cite{Banerjee2013,Kolliopoulos98approximatingdisjoint-path,Ozdaglar2003,Baveja00approximationalgorithms,Klinkowski2016} bin-packing based approaches~\cite{skorin07}, Monte Carlo search~\cite{Pham2012}, post-optimization~\cite{post14} and  greedy algorithms~\cite{chen96,sumpter2013, srinivas03, Manohar2002}, a large number of heuristics have also been used to obtain approximate solutions in both problems, among them genetic algorithms~\cite{Noronha10abiased,Storn1997,Hsu2015}, ant colony optimization~\cite{ blesa04}  and particle swarm optimization~\cite{Hsu2017}. Specifically in the area of optical communication networks, it is common to use integer/linear programming and its variants for small networks, to obtain exact results, and a variety of heuristics for larger networks.  In practice, current optical networks use overprovision of capacity to compensate for sub-optimal routing, resulting in both over-engineering and underutilized capacity~\cite{Agrell_2016, Bayvel2016}. 
	
	The main challenge we address here is the RWA under heavy traffic using \emph{multiple wavelengths} and a very large number of origin-destination pairs under the NDP and EDP constraints and for various objective functions. Globally optimal routing is a non-localized difficult problem but adding the NDP/EDP restrictions and having a large number of different wavelengths increases the complexity considerably, making the problem intractable for existing algorithms~\cite{de2014shortest,altarelli2015edge}. We map the globally-optimal routing problem in the presence of multiple wavelengths onto a multi-layer replica of the original graph and utilize probabilistic optimization approaches. The methods developed here are based on message passing techniques, developed independently in  several fields including statistical physics, computer science and information theory~\cite{sg87,Pearl82,Gallager63} but are closely interlinked~\cite{inf09,Kabashima_1998}. The methods we develop allow for messages, in the form of conditional probability values to be passed between \emph{nodes and the replicated networks representing the different wavelengths}, in a way that keeps the algorithms scalable and applicable even for a large number of wavelengths, transmissions (corresponding to source-destination pairs) and nodes.  
	
	The main result of this paper is the derivation of principled scalable algorithms, capable of obtaining approximate solutions for routing problems in large graphs, where the number of transmissions is of similar order  to that of the number of free variables (vertices/edges) and a large number of wavelengths, under the NDP and EDP constraints and for various objective functions, both convex and concave. The algorithm also accommodate cases where the number of transmissions is much larger than the number of vertices (quadratic with respect to the number of vertices). The computational complexity of the NDP/EDP algorithms for sparse graphs is ${O(MQ(M+N+Q))/}O(MQ(M/N+N+Q))$, with $N$ the number of vertices, $M$ the number of transmissions and $Q$ the number of wavelengths. The algorithm has been tested for a variety of sparse network topologies, both synthetic random graphs and real optical communication networks, and for different objective functions, showing excellent results in obtaining high quality approximate solutions. Among the generic networks examined are random regular graphs, Erd\H{o}s R\'{e}nyi (ER)~\cite{erdos59a} and scale free networks~\cite{Onnela_2007}, while the realistic networks considered include the British 22 nodes (BT22) and US 60 node (CONUS)~\cite{conus60}backbone optical communication networks. When tested on small networks against known results obtainable using unscalable methods like variants of integer/linear programming, it was shown to provide the optimal routing results.
	
	The results provide the maximal number of communication pairs that could be accommodated given the network size, topology and number of wavelengths used; the minimal number of wavelengths required for a given network, topology and communication pairs; and the resulting utilization of edges. They identify the impact of using the suggested algorithm on the average path length, the utilization of wavelengths per edge and how it can be controlled using concave and convex objective functions.
	
	In addition, our algorithms can be used for routing transmissions across networks in single instances, study the limitations of heterogeneous networks of different degree distributions, with variable edge signal-to-noise ratios and wavelength availability. Moreover, our algorithms could be employed in the design of new infrastructure, especially through the use of concave cost functions that consolidate routes, by determining the least important routes that could be removed with little effect on the network throughput or resilience. These are of both academic and practical values since the performance of optical communication networks is often directly related to their capacity limits, traffic congestion, rate of information flow and bandwidth flexibility. Moreover, we also studied a switching model, where wavelength can be converted (switched)  at the vertices (transceivers) to settle an open and much debated question on the merit of wavelength-converters~\cite{Sato2013,Gerstel2012,Teipen2012} for increasing throughput, resource utilization and resilience in optical communication networks.
	
	While we mainly focus here on the optical communication network application and test the efficacy of the method on networks and number of wavelengths that are relevant to this application domain, one should point out that these problems are highly relevant to other domains. For instance, multi-wavelength NDP/EDP  are relevant to both 5G and the future 6G wireless communication systems  and wireless ad-hoc communication networks in the relay setting, where each node can act as a relay. Our algorithms can reduce path overlaps, which represent transmissions in similar wavelength, resulting in signal interference and low transmission quality, or to consolidate paths since longer paths result in signal degradation and the need for higher transmission power~\cite{sumpter2013,srinivas03,adhoc}. Another application is the design of very large system multilayer integrated circuits (VLSI), where  non-overlapping wired paths to connect different components are sought to avoid cross-path interference. In all cases, higher throughput, robustness and lower latency can be achieved for the same resource by obtaining a good approximation to the globally optimal solution. Practical algorithms for various applications often depend on the specific network topologies considered~\cite{mesh96} and are aimed at maximizing the number of paths routed~\cite{mndp}.

	The reminder of the paper is organized as follows: in Sec.~\ref{sec:model} we  introduce the model used followed by the message-passing based algorithmic solutions for NDP, EDP and wavelength switching scenarios in Sec.~\ref{sec:algorithms}. Results obtained from numerical studies on a range of synthetic and real networks and a variety of objective functions are presented in Sec.~\ref{sec:results} followed by a discussion on their computational complexity. Possible extensions of the framework to accommodate real-world scenarios such as edges with different signal-to-noise ratios or wavelength availability are presented in Sec.~\ref{sec:extensions}. Finally, we discuss the efficacy of the methods developed and point to future research directions in Sec.~\ref{sec:conclusion}.

\section{Model\label{sec:model}}

We consider a dense wavelength-division multiplexing (DWDM) optical network 
$G(V,E)$, with $V\equiv \{i ~\vert~ i\in G\}$ and $|V|=N$, the set of nodes representing transceivers and $E\equiv \{(i,j) ~\vert~ (i,j)\in G\}$ the set of edges, such that the indices $(i,j)$ represent the optical fiber between node $i$ and $j$. For a network which uses $Q$ wavelength channels to deliver $M$ transmissions, we introduce a variable $s_{i,j}$  on the link from node $i$ to $j$ such that $s^a_{i,j}=  s$ or $-s$ if transmission $s$ passes from node $i$ to $j$ or from $j$ to $i$ respectively, through link $(i,j)$ using wavelength $a$; the transmission $s$ corresponds to one of the origin-destination pair $\{0,1,\dots,M$\}. A similar variable $s^a_i$ is defined for node $i$ in the case of NDP, as explained in Sec.~\ref{sec:algorithms}. Since each transmission has to occupy an individual wavelength channel on a link, more wavelengths are generally required for more transmissions, i.e. a larger value of $Q$ is required for a larger $M$. For specific network instances with $Q$ wavelength channels, there exists a maximum number of transmissions denoted as $M_{\rm max}$ which can be transmitted; alternatively, one can define the minimum number of wavelength channels, i.e. $Q_{\rm min}$, which accommodate all $M$ transmissions on a specific instance. The relationship between $M_{\rm max}$ and $Q$, or between {$M$ and} $Q_{\rm min}$, would be highly relevant for characterizing the maximum capacity of optical networks.

This framework can accommodate a variety of objective functions; here, we consider the sum of the cost (or utility) on each link to be the objective function for optimization, given by
\begin{equation}
\label{eq:hamilton}
H(\vec{\vec s})= \sum_{(i,j)} F_{i,j} \del[3]{Q-\sum_{a=1}^Q \delta_{s_{i,j}^a}^0},
\end{equation}
where $\delta^y_x$ is the Kronecker delta such that $\delta^y_x=1$ if $x=y$ and $\delta^y_x=0$ otherwise; the function $F_{i,j}$ denotes the cost on link $(i,j)$ as a function of the argument in parenthesis; while it can take an arbitrary form we will mostly focus on simple polynomial functions. In the context of statistical physics, we introduce the inverse temperature $\beta$, and the partition function $Z$ of the system is given by
\begin{equation}
\label{eq:partition}
\begin{aligned}
Z(\beta)= &~ \sum_{\vec{\vec s}}\Omega(\vec{\vec s})\ee^{-\beta H(\vec{\vec s})},\\¬
=&~ \sum_{\vec{\vec s}} \Omega(\vec{\vec s})\prod_{(i,j)} \ee^{-\beta F_{i,j}(\vec s_{i,j})},
\end{aligned}
\end{equation}
where $\Omega(\vec{\vec s})$ is an indicator function such that $\Omega(\vec{\vec s})=1$ if $\vec{\vec s}$ satisfies all the constraints of the problem or otherwise $\Omega(\vec{\vec s})=0$. The double vector notation comes to emphasize dependence on both topology and wavelength.

\begin{figure*}[hbt!]
\centering
\subfigure[Node-disjoint]{\includegraphics[width=0.32\textwidth]{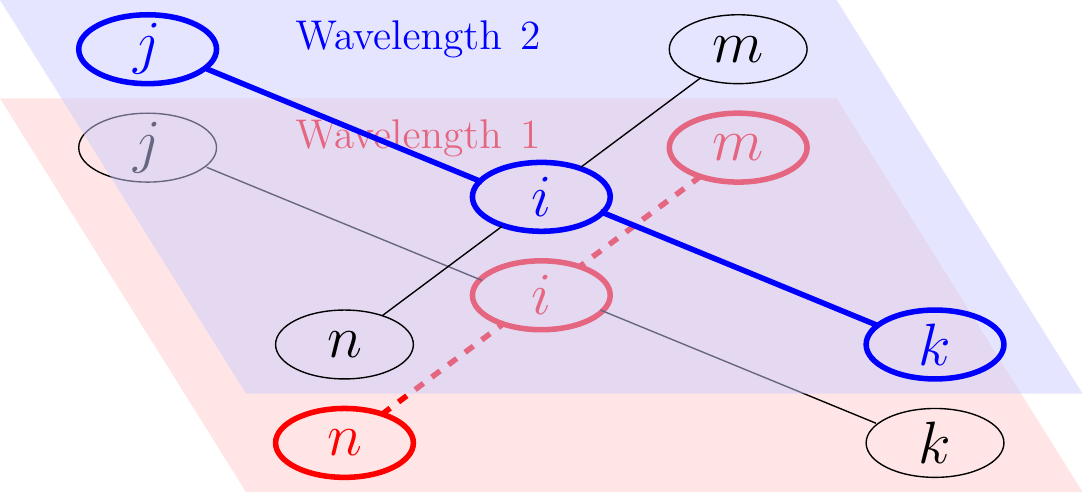}
\label{fig:nodeDisjointDemo}}
\subfigure[Node-disjoint wavelength-switching]{\includegraphics[width=0.32\textwidth]{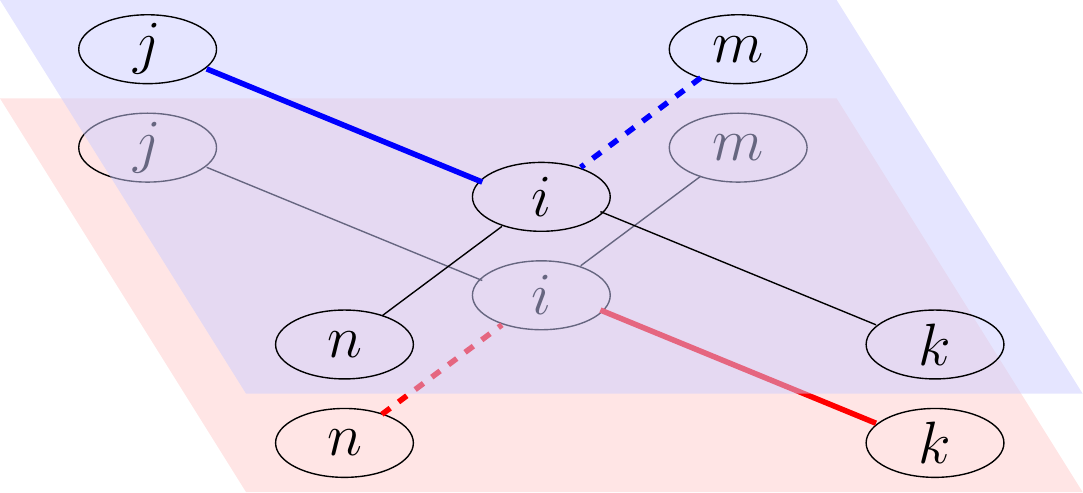}
\label{fig:nodeSwitchingDemo}}
\subfigure[Edge-disjoint]{\includegraphics[width=0.32\textwidth]{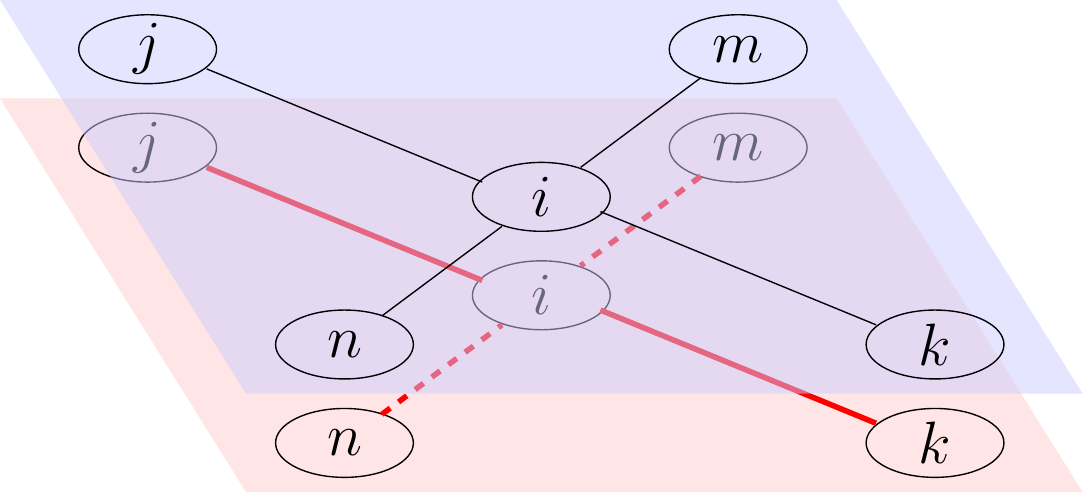}
\label{fig:edgeDisjointDemo}}
\caption{An exemplar graph with $N=5$ nodes and $4$ edges, where $M=2$ transmissions with origins and destinations $j$ and $k$ (solid), and $m$ and $n$ (dashed) respectively, are transmitted by $Q= 2$ wavelength channels represented by the red and blue layers. (a) The node-disjoint (NDP) case, where the whole transmission path uses the same wavelength. For instance, the red wavelength channel of node $i$ is used by the transmission from $m\to n$, so the red node $i$ cannot be a part of the other transmission from $j\to k$, which instead uses the blue wavelength channel. (b) The wavelength-switching scenario of NDP {(WS)}, where the two transmission switch their wavelength channels at node $i$; both (a) and (b) are valid under this switching scenario. (c) The edge-disjoint (EDP) case, where the red layer is sufficient for accommodating  the two transmissions, and the blue layer is idle.}
\end{figure*}

We now summarize the constraints of the optimization problem. Firstly, the route for each transmission must be contiguous. A loopless path is a sequence of non-repeating nodes from origin to destination, for example, path $s$ is constructed as $O_s \to \cdots \to j\to i \to k \to \cdots \to D_s$, where $O_s$ and $D_s$ are the origin and destination pair of transmission $s$. Secondly, for an intermediate node $i$ along the path there exist only two used edges $(j,i)$ and $(i,k)$ for that wavelength and transmission. This constraint could be expressed as follows: if $s^a_{i,j}=s\neq 0$, then 
\begin{equation}
\sum_{k\in\partial i\setminus j} (1-\delta^0_{s^a_{i,k}})=1 \mbox{ and } \sum_{k\in\partial i\setminus j}s^a_{i,k}= -s,
\end{equation}
where $\partial i \equiv \{j ~\vert~ (i,j)\in E\}$ is the set of the nearest  neighbors of node $i$, and $\partial i \setminus j \equiv \partial i - \{j\}$ is the subset of $\partial i$ except node $j$.
Regarding the wavelength constraints along the path, we will consider three scenarios in the subsequent analyses: 
\begin{enumerate}
\item
\emph{Node-disjoint paths} (NDP) --- where only a single transmission is allowed to utilize a specific wavelength channel on a node~\cite{de2014shortest}, but there can be multiple transmissions using  different channels through the same node as shown in Fig.~\ref{fig:nodeDisjointDemo}; this may correspond to transceivers which can only process a single transmission for each individual wavelength. The expression of the node-disjoint constraint is $s_{i,j}^a=s \neq 0 \Rightarrow \forall k\in\partial i\setminus j: s_{i,k}^a\in \{0, -s\}$. 
\item
\emph{Wavelength-switching} {(WS)} with NDP --- where all incoming or outgoing transmissions to transceivers (nodes) use different wavelength channels, but transmissions are allowed to switch between wavelength channels at the transceivers as shown in Fig.~\ref{fig:nodeSwitchingDemo}, leading to a larger routing flexibility. 
\item
\emph{Edge-disjoint paths} (EDP) --- where multiple transmissions using the same wavelength channel, are allowed to be routed through any given node~\cite{altarelli2015edge}but cannot share an edge, as shown in Fig.~\ref{fig:edgeDisjointDemo}. This is the typical scenario in optical communication networks.
\end{enumerate}

\section{Message-passing algorithms\label{sec:algorithms}}

To derive optimization algorithms to allocate simultaneously the optimal route and wavelength for a large number of transmissions, we solve the problem on multi-layer graphs where each layer represents a different wavelength, and messages are passed within each layer for assignment of routes and between layers for allocation of wavelength. By applying the cavity approach from the study of spin glass systems~\cite{sg87,cavity}, we can derive distributed message-passing algorithms for optimizing transmission routes in optical networks for NDP, WS and EDP scenarios.

\subsection{Node-disjoint routing}

First, we consider the NDP routing scenario~\cite{de2014shortest} shown in Fig.~\ref{fig:nodeDisjointDemo}. The network is represented by a factor graph with two types of variables, namely $s^a_i$ defined on nodes and $s^a_{i,j}$ defined on links. According to the Bethe-Peierls approximation~\cite{BethePeierls,mezard2001bethe}, we assume only large loops exist in the network such that all neighboring nodes and edges of a node $i$ are nearly independent in the absence of $i$. 

To derive the \emph{multi-wavelength routing} algorithm, we first define $p_{i\to j}$ to be the message from node $i$ to edge $(i,j)$ and $q_{i\to j}$ to be the message from edge $(i,j)$ to node $j$. Both messages are conditional probabilities: $p_{i\to j}(s)$ is the probability of edge $(i,j)$ to be in state $s$ due to the state of node $i$ and $q_{i\to j}(s)$ the probability of edge $(i,j)$ to be in state $s$ without the interaction from node $i$. We then write a closed set of recursion relations to express the message $p_{i\to j}$ in terms of $q_{k\to i}$, as well as another set of relations representing $q_{i\to j}$ in terms of $p_{i\to j}$ and $p_{j\to i}$:
\begin{equation}\small
\label{eq:belief-propagation}
\left\{
\begin{aligned}
p_{i\to j}^a(0) \sim &~ \prod_{k\in\partial i\setminus j} q_{k\to i}^a(0)+ \\
&~ \sum_{m,n\in\partial i\setminus j\atop s\neq 0} q_{m\to i}^a(s)q_{n\to i}^a(-s)\prod_{k\in\partial i\setminus j,m,n} q_{k\to i}^a(0),\\
p_{i\to j}^a(s) \sim &~ \sum_{k\in\partial i\setminus j} q_{k\to i}^a(s) \prod_{l\in\partial i\setminus j,k} q_{l\to i}^a(0);\\
q_{i\to j}^a(0) \sim &~ p_{i\to j}^a(0)  \sum_n \ee^{-\beta F_{i,j}(n)} \sum_{b\neq a:\atop s^b=0,1}\delta^n_{\sum\limits_{b\neq a}s^b}\prod_{b\neq a} \tilde q^b_{i,j}(s^b),\\
q_{i\to j}^a(s) \sim &~ p_{i\to j}^a(s) \sum_n \ee^{-\beta F_{i,j}(n+ 1)} \sum_{b\neq a:\atop s^b=0,1}\delta^n_{\sum\limits_{b\neq a}s^b}\prod_{b\neq a} \tilde q^b_{i,j}(s^b),
\end{aligned}
\right.
\end{equation}
where we have further defined the auxiliary quantities $\tilde q^b_{i,j}$ given by 
\begin{equation}
\left\{
\begin{aligned}
\tilde q_{i,j}^a(0) =&~ \hat q_{i,j}^a(0),\\
\tilde q_{i,j}^a(1) =&~ \sum_{s\neq 0} \hat q_{i,j}^a(s);\\
\hat q_{i,j}^a(s)=&~ p_{i\to j}^a(s) ~p_{j\to i}^a(-s).
\end{aligned}
\right.
\end{equation}

\begin{figure}[t!]
\centering
\includegraphics[width=0.48\textwidth]{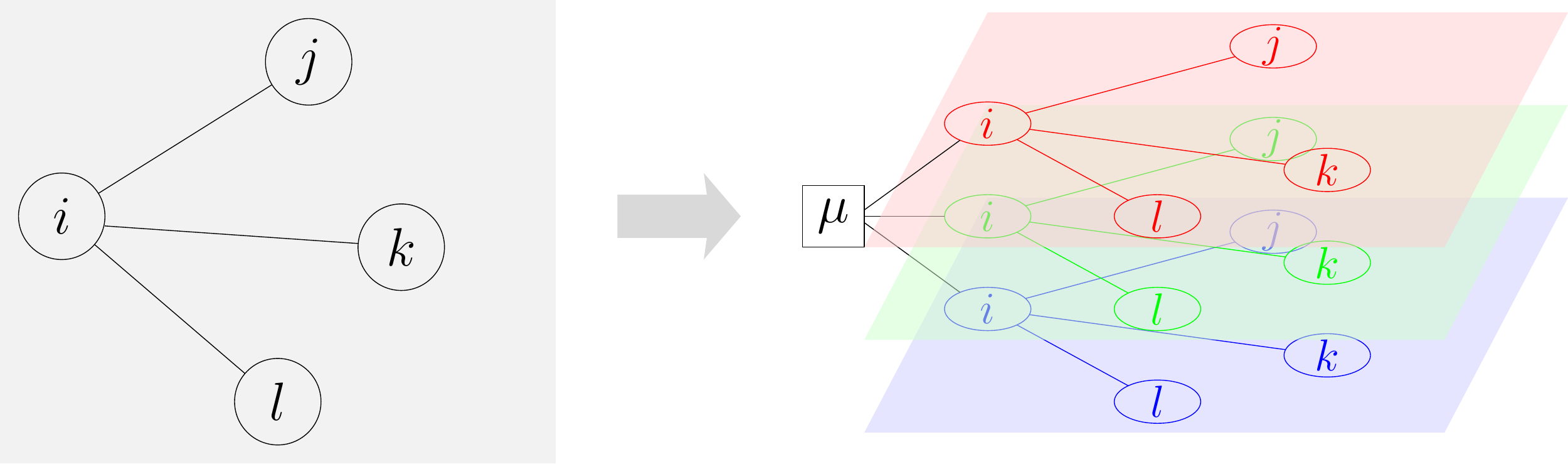}
\caption{\label{fig:ExtraNodeDemo}Mapping the original network (left) onto multi-layer replica networks that use different wavelengths (right). In this example, node $i$ is the origin/destination of transmission $|\mu|$. Introducing an auxiliary node $\mu$, denoted by a square and connected to nodes $i$ at each of the layers, facilitates message passing between the new node and the different layers to determine the allocation of transmissions to  wavelengths. These auxiliary nodes also facilitate the interaction among different wavelengths.}
\end{figure}

For brevity we omit the normalization term and use the notation ``$\sim$'' instead of the equality symbol. For a brief explanation of how these equations are construed: the first  equation of~\eqref{eq:belief-propagation},  looks at the probability for no transmission on the edge $i\rightarrow j$, as a summation of the probability of no transmission entering $i$ (first term) and the probability of transmissions arriving at $i$ from node $m$ but leaving through some other node $n$ (second term); the second equation looks at the message $s$ arriving at node $i$ but not leaving through any other edge but $i\rightarrow j$; the third and fourth equations describe the probability of edge $i\rightarrow j$ to be in a given state $0$ or $s$, given the related cost on the edge, in conjunction with all other wavelengths. In Eq.~\eqref{eq:belief-propagation}, node messages using wavelength $a$ depend only on the messages from their neighbors using the same wavelength and the network is effectively mapped to a system with $Q$ separate layers, each of  which employs a different wavelength channel, as shown in Fig.~\ref{fig:nodeDisjointDemo}; the interdependence between wavelengths is considered at the origin and destination of individual transmissions (Fig.~\ref{fig:ExtraNodeDemo}), as discussed below. 

With the messages $p_{i\to j}$ and $q_{i\to j}$ having converged to stable values, one can express the marginal probability of node $i$ being in state $s$ using wavelength $a$ as
\begin{equation}
\left\{
\begin{aligned}
p_i^a(0) {\sim}&~ \prod_{j\in\partial i} q_{j\to i}^a(0),\\
p_i^a(s) {\sim}&~ \sum_{m,n\in\partial i} q_{m\to i}^a(s) q_{n\to i}^a(-s) \prod_{j\in\partial i\setminus m,n} q_{j\to i}^a(0).
\end{aligned}
\right.
\end{equation}
The marginal probability of edge $(i,j)$ being in state $s$ with wavelength $a$ is given by 
\begin{equation}\small
\left\{
\begin{aligned}
q_{i,j}^a(0) \sim&~ \hat q_{i,j}^a(0) \cdot \sum_{n}\ee^{-\beta F_{i,j}(n)} \sum_{b\neq a:\atop s^b=0,1} \delta_{\sum\limits_{b\neq a}s^b}^n \prod_{b\neq a} \tilde q^b_{i,j}(s^b),\\
q_{i,j}^a(s) \sim&~ \hat q_{i,j}^a(s)\cdot \sum_{n}\ee^{-\beta F_{i,j}(n+ 1)} \sum_{b\neq a:\atop s^b=0,1} \delta_{\sum\limits_{b\neq a}s^b}^n \prod_{b\neq a} \tilde q^b_{i,j}(s^b),
\end{aligned}
\right.
\end{equation}
where the factors after the dot symbols correspond to contributions from all neighboring links of $(i, j)$ to the partition function, which takes into account all possible variable configurations in the trees terminated at the neighboring edges.

To simplify the algorithms, we introduce the variables {$\phi\equiv -\frac{1}{\beta}\log q$ and $\psi\equiv -\frac{1}{\beta}\log p$}, and take the zero-temperature limit for optimization  $\beta\to \infty$. The  recursion relations of messages in Eq.~\eqref{eq:belief-propagation} thus become equivalent to the min-sum belief propagation relations, given by
\begin{equation}
\label{eq:ND_mess}
\left\{
\begin{aligned}
\psi^a_{i\to j}(0) \sim &~ \min\bigg\{\sum_{k\in\partial i\setminus j} \phi^a_{k\to i}(0), ~\min_{m,n\in\partial i\setminus j\atop s\neq 0} \Big[\phi^a_{m\to i}(s)\\
&~ \quad +\phi^a_{n\to i}(-s)+ \sum_{k\in\partial i\setminus j,m,n} \phi^a_{k\to i}(0)\Big]\bigg\},\\
\psi^a_{i\to j}(s) \sim &~ \min_{k\in\partial i\setminus j}\Big[\phi_{k\to i}^a(s)+ \sum_{l\in\partial i\setminus j,k}\phi^a_{l\to i}(0)\Big];\\
\phi^a_{i\to j}(0) \sim &~ \psi^a_{i\to j}(0)+\\
&~ \min_n \Big[F_{i,j}(n)+ \min_{\sum\limits_{b\neq a}s^b = n:\atop s^b=0,1} \sum_{b\neq a} \tilde \phi_{i,j}^b(s^b)\Big],\\
\phi^a_{i\to j}(s) \sim &~ \psi^a_{i\to j}(s)+\\
&~  \min_n \Big[F_{i,j}(n+ 1)+ \min_{\sum\limits_{b\neq a}s^b = n:\atop s^b=0,1} \sum_{b\neq a} \tilde \phi_{i,j}^b(s^b)\Big],
\end{aligned}
\right.
\end{equation}
where
\begin{equation}
\label{eq:ND_tilde_phi}
\left\{
\begin{aligned}
\tilde \phi_{i,j}^a(0)=&~ \psi^a_{i\to j}(0) + \psi^a_{j\to i}(0),\\
\tilde \phi_{i,j}^a(1)=&~ \min_{s\neq 0} \Big[\psi_{i\to j}^a(s)+ \psi_{j\to i}^a(-s)\Big].
\end{aligned}
\right.
\end{equation}

For each individual transmission, we then introduce auxiliary nodes labeled as $\mu= \pm \vert\mu\vert$, which connect to the origin and destination nodes in each of the $Q$ wavelength network layers, respectively (see in Fig.~\ref{fig:ExtraNodeDemo}). These auxiliary origin-destination pairs of each transmission communicate with all wavelength replica networks to determine the path and wavelength channel allocated to each transmission, using the message passing algorithm, given by 
\begin{equation}
\label{eq:ND_endpoint}
\left\{
\begin{aligned}
\phi_{\mu\to a}(0) \sim&~ \min_{b\neq a} \sbr[2]{\phi_{b\to \mu}(-\mu)+ \sum_{c\neq a, b} \phi_{c\to \mu}(0)},\\
\phi_{\mu\to a}(\mu) \sim &~ 1+\sum_{b\neq a}\phi_{b\to \mu}(0),\\
\phi_{\mu\to a}(s) \sim &~ \infty, \quad  s\neq 0,\mu,
\end{aligned}
\right.
\end{equation}
such that the transmission routes are determined independently on each wavelength channel. After introducing the auxiliary nodes, we treat the messages to and from them as to other network nodes. For example, the neighboring nodes set of node $i$ in Fig.~\ref{fig:ExtraNodeDemo} is $\partial i= \{j,k,l,\mu\}$, and the calculation of messages $i\to j$ and $i\to \mu$ both follow Eq.~\eqref{eq:ND_mess}. The marginal probability of edge $(i,j)$ being in state $s$ using wavelength $a$ is then given by
\begin{equation}
\label{eq:ND_marginal}
\left\{
\begin{aligned}
\phi^a_{i,j}(0) \sim &~ \psi_{i\to j}^a(0)+ \psi_{j\to i}^a(0)+ \min_n \Big[F_{i,j}(n)+ \\
&~ \quad\min_{\sum\limits_{b\neq a}s^b = n:\atop s^b=0,1} \sum_{b\neq a} \tilde \phi_{i,j}^b(s^b)\Big],\\
\phi^a_{i,j}(s) \sim &~ \psi_{i\to j}^a(s)+ \psi_{j\to i}^a(-s)+ \min_n \Big[F_{i,j}(n+ 1) \\
&~ \quad+\min_{\sum\limits_{b\neq a}s^b = n:\atop s^b=0,1} \sum_{b\neq a} \tilde \phi_{i,j}^b(s^b)\Big],
\end{aligned}
\right.
\end{equation}
and the state $s_{i,j}^a$ of wavelength $a$ of edge $(i,j)$ is determined by
\begin{equation}
\label{eq:ND_state_chooser}
s_{i,j}^a= \arg\min_{s} \phi_{i,j}^a(s),
\end{equation}
which ultimately leads to the optimized configuration of routes for all $M$ transmissions through the optical network with $Q$ wavelength channels.

\subsubsection{Linear cost\label{sec:ND_short}}

While the objective function can take many different forms we use a simple power as the cost function $F_{i,j}(x)$ on edges $x^\gamma$~\cite{yeung2012competition,badiu2021self}, as it provides a good example of both concave and convex costs. When $\gamma= 1$, the cost function is linear and equivalent to the total length of all transmissions $L\equiv \sum_{(i,j);a} \del[1]{1-\delta^0_{s_{i,j}^a}}$, and optimizing it is equivalent to finding the shortest average path.

In this case ($\gamma=1$), the message-passing equations Eq.~\eqref{eq:ND_mess} can be simplified to
\begin{equation}
\label{eq:ND_mess_shortest}
\left\{
\begin{aligned}
\phi_{i\to j}^a(0) \sim &~ \min\bigg\{\sum_{k\in\partial i\setminus j} \phi^a_{k\to j}(0), \min_{m,n\in\partial i\setminus j\atop s\neq 0} \Big[\phi^a_{m\to i}(s)+\\
&~\quad\phi^a_{n\to i}(-s)+\sum_{k\in\partial i\setminus j,m,n}\phi^a_{k\to i}(0)\Big]\bigg\},\\
\phi_{i\to j}^a(s) \sim&~ 1+ \min_{k\in\partial i\setminus j}\Big[\phi_{k\to i}^a(s)+\sum_{l\in\partial i\setminus j,k}\phi^a_{l\to i}(0)\Big].
\end{aligned}
\right.
\end{equation}
where and the variable $\psi$ in Eq.~\eqref{eq:ND_mess} can be omitted. The marginal quantities are then given by
\begin{equation}
\label{eq:ND_marginal_shortest}
\phi^a_{i,j}(s)= \phi_{i\to j}^a(s)+ \phi_{j\to i}^a(-s)+ (\delta_{s}^0- 1),
\end{equation}
and the messages to the auxiliary nodes at the origins and destinations are the same as Eq.~\eqref{eq:ND_endpoint}.

\subsubsection{Switching wavelength channels at nodes}

A generalization of the  NDP scenario is to allow for transmissions to change  wavelengths at nodes {(WS)} as shown in Fig.~\ref{fig:nodeSwitchingDemo}, which is equivalent to a single-wavelength network routing with both node and edge capacity being $Q$, and utilize other multiplexing techniques than wavelength division (e.g. code or time division multiplexing). A variable $\tau$ is introduced to enforce this capacity constraint on nodes. In this case, the message-passing equations are given by
\begin{equation}
\label{eq:NS_mess}
\left\{
\begin{aligned}
\phi_{i\to j}^\mu(0) \sim &~ \min\bigg\{\tau^\mu_i(0)+ \sum_{k\in\partial i\setminus j}\phi_{k\to i}^\mu(0), ~\tau_i^\mu(1)+\\
&~  \min_{m,n\in\partial i\setminus j}\Big[\phi_{m\to i}^\mu(1)+ \phi_{n\to i}^\mu(-1)\\
&~ +\sum_{k\in\partial i\setminus j,m,n}\phi_{k\to i}^\mu(0)\Big]\bigg\},\\
\phi_{i\to j}^\mu(\pm 1) \sim &~ \tau_i^\mu(1)+ 1+\\
&~ \min_{k\in\partial i\setminus j}\Big[\phi_{k\to i}^\mu(\pm 1)+ \sum_{l\in\partial i\setminus j,k} \phi_{l\to i}^\mu(0)\Big],
\end{aligned}
\right.
\end{equation}
where $\tau_i^\mu$ denotes the summation over $\tilde \psi_i$ from a total of at most $Q$ transmissions, excluding $\mu$, that pass through node $i$, given by
\begin{equation}
\label{eq:NS_capacity_constraint} 
\left\{
\begin{aligned}
\tau_i^\mu(0) =&~ \min_{\sum_{\nu\neq \mu}\sigma^\nu\leq Q_i\atop\nu\neq \mu: \sigma^\nu=0,1}  \sum_{\nu\neq \mu} \tilde \psi_i^\nu(\sigma^\nu),\\
\tau_i^\mu(1) =&~ \min_{1+\sum_{\nu\neq \mu}\sigma^\nu\leq Q_i\atop\nu\neq \mu: \sigma^\nu=0,1}  \sum_{\nu\neq \mu} \tilde \psi_i^\nu(\sigma^\nu).
\end{aligned}
\right.
\end{equation}
Specifically, $\tau_i^\mu(1)$ could be understood as indicating that there are free wavelength channels on $i$ that $\mu$ could take, and $\tau_i^\mu(0)$ indicates that transmission $\mu$ does not pass through node $i$.
The quantity $\tilde \psi_i^\mu$ is an auxiliary variable related to the probability that node $i$ would be a part of the path of transmission $\mu$, which leads to
\begin{equation}
\left\{
\begin{aligned}
\tilde \psi_i^\mu(0) \sim &~ \sum_{j\in\partial i} \phi_{j\to i}^\mu(0),\\
\tilde \psi_i^\mu(1) \sim &~ \min_{j,k\in\partial i}\bigg[ \phi_{j\to i}^\mu(1) +\phi_{k\to i}^\mu(-1)+ \sum_{l\in\partial i\setminus j,k}\phi_{l\to i}^\mu(0)\bigg].
\end{aligned}
\right.
\end{equation}
The messages from the origin and the destination of transmission $\mu$ to the neighboring edges are slightly different from the previous case of NDP without switching. {In Eq.~\eqref{eq:NS_capacity_constraint} $Q_i= Q$, and  $\tilde \psi_i(1)= 0$ and $\tilde \psi(0)=\infty$ for the transmissions using node $i$ as their origin or destination. Alternatively, one can set $Q_i= Q- M_i$, where $M_i$ is the number of transmissions using node $i$ as their origin or destination, and then exclude these transmissions when one calculates $\tau_i$.} Here, we first introduce a variable {$\mu_i= \pm 1$ to denote node $i$ as the origin or destination of transmission $\mu$ respectively}, given by
\begin{equation}
\left\{
\begin{aligned}
\phi_{i\to j}^\mu(0) \sim &~ \min_{k\in\partial i\setminus j} \Big[\phi_{k\to i}^\mu(-\mu_i)+ \sum_{l\in\partial i\setminus j,k}\phi_{k\to i}^\mu(0)\Big],\\
\phi_{i\to j}^\mu(\mu_i)\sim &~ 1+ \sum_{k\in\partial i\setminus j} \phi_{k\to i}^\mu(0),\\
\phi_{i\to j}^\mu(-\mu_i)= &~ \infty.
\end{aligned}
\right.
\end{equation}

The variable $\phi_{i,j}^\mu$ determines the state of edge $(i,j)$, i.e. whether it is a part of the path of transmission $\mu$ or not, and its expression is given by
\begin{equation}
\phi_{i,j}^\mu(\sigma)= \phi_{i\to j}^\mu(\sigma)+ \phi_{j\to i}^\mu(-\sigma)+ (\delta_\sigma^0-1),
\end{equation}
such that the optimized state is $\sigma_{i,j}^\mu= \arg\min_{\sigma} \phi_{i,j}^\mu(\sigma)$.

\subsection{Edge-disjoint routing}

Edge-disjoint path routing (EDP) is similar to NDP routing but is less restrictive, since nodes can accommodate any number of paths with the same wavelength but edges do not~\cite{altarelli2015edge} (Fig.~\ref{fig:edgeDisjointDemo}). In other words, for node $i$, if there exists an edge $(i,j)$ with $s_{i,j}^a= s_0\neq 0$, then there exists one and only one edge $(i,k)$ with $s_{i,k}^a= -s_0$ continuity of transmission path,  while other neighboring edges using \emph{the same wavelength channel} $a$ could be either in state $0$ or take up other transmissions such that $s_{i,m_1}^a=-s_{i,n_1}^a= s_1\neq 0$, $s_{i,m_2}^a=-s_{i,n_2}^a= s_2\neq 0$ etc. In comparison, for NDP at most two variables $s_{i,j}^a$ for neighboring nodes $j\in\partial i$ can assume a non-zero value. { In other words, single-wavelength EDP is equivalent to a generalized version of WS, where the capacity of each node is non-uniform and determined by its degree and the number of it being chosen as origins or destinations of transmissions.} Consequently, the message passing equations for EDP scenarios are more complicated and their computational complexity is higher. The corresponding message passing equations are given by 
\begin{equation}\small
\label{eq:ED_mess}
\left\{
\begin{aligned}
\psi_{i\to j}^a(0) \sim &~ \min_{\mbox{matched}\atop\mbox{pairs: } \vec s_{\partial i\setminus j}} \sum_{k\in\partial i\setminus j} \phi_{k\to i}^a(s_k),\\
\psi_{i\to j}^a(s) \sim &~ \min_{k\in\partial i\setminus j} \Big[\phi_{k\to i}^a(s)+ \min_{\mbox{matched}\atop\mbox{pairs: }\vec s_{\partial i\setminus j,k}} \sum_{l\in\partial i\setminus j,k} \phi_{l\to i}^a(s_l)\Big];\\
\phi_{i\to j}^a(0) \sim &~ \psi_{i\to j}^a(0)+ \min_n\Big[F_{i,j}(n)+ \min_{\sum_{b\neq a}s^b=n:\atop s^b=0,1} \sum_{b\neq a} \tilde \phi_{i,j}^b(s^b)\Big],\\
\phi_{i\to j}^a(s) \sim &~ \psi_{i\to j}^a(s)+ \min_n \Big[F_{i,j}(n+1)+\\
&\qquad ~\min_{\sum_{b\neq a} s^b=n:\atop s^b=0,1} \sum_{b\neq a} \tilde \phi_{i,j}^b(s^b)\Big],
\end{aligned}
\right.
\end{equation}
where the message from node $i$ to edge $(i,j)$ (first two equations) are different from those of NDP in Eq.~\eqref{eq:ND_mess}, while Eq.~\eqref{eq:ND_tilde_phi}--\eqref{eq:ND_state_chooser} in the NDP scenario apply also here.

It is numerically difficult to compute the message passing relations in Eq.~\eqref{eq:ED_mess}, since  all feasible configurations satisfying the constraints have to be considered by the first two equations, {which results in a computational complexity of
\begin{equation}
\sum_{n=0}^{\min\{\lfloor K/2\rfloor, M\}} {}^MC_n ~{}^KP_{2n},
\end{equation}
where $K= \vert\partial i\vert-1$ is the number of neighboring edges  of node $i$, $M$ is the total number of transmissions, $\lfloor x\rfloor$ is the floor function that is equal to the greatest integer less than or equal to $x$, ${}^nP_k=\frac{n!}{(n-k)!}$ is the number of ordered permutations of $k$ out of $n$ elements and ${}^nC_k=\frac{n!}{k!(n-k)!}$ the number of unordered combinations.} To simplify the computation we map the task on to the maximum weight matching problem~\cite{lovasz2009matching,altarelli2015edge}. As in~\cite{altarelli2015edge}, we consider all the possible pairs of considered edges $\{(i,k)~\vert~ k\in\partial i\setminus j\}$ and obtain the weight of each pair 
\begin{equation}
\label{eq:ED_weights}
w_{k,l}= -\min_{s=-M}^M \big[\phi_{k\to i}(s)+ \phi_{l\to i}(-s)\big].
\end{equation}
Then, we construct a weighted graph with $\vert\partial i\vert-1$ nodes where the weight of each edge is given by Eq.~\eqref{eq:ED_weights}. After that, $\psi_{i\to j}^a$ in Eq.~\eqref{eq:ED_mess} could be obtained by the maximum weight matching algorithm~\cite{galil1986ev,altarelli2015edge}.

\begin{figure}[t!]
\centering
\includegraphics[width=0.4\textwidth]{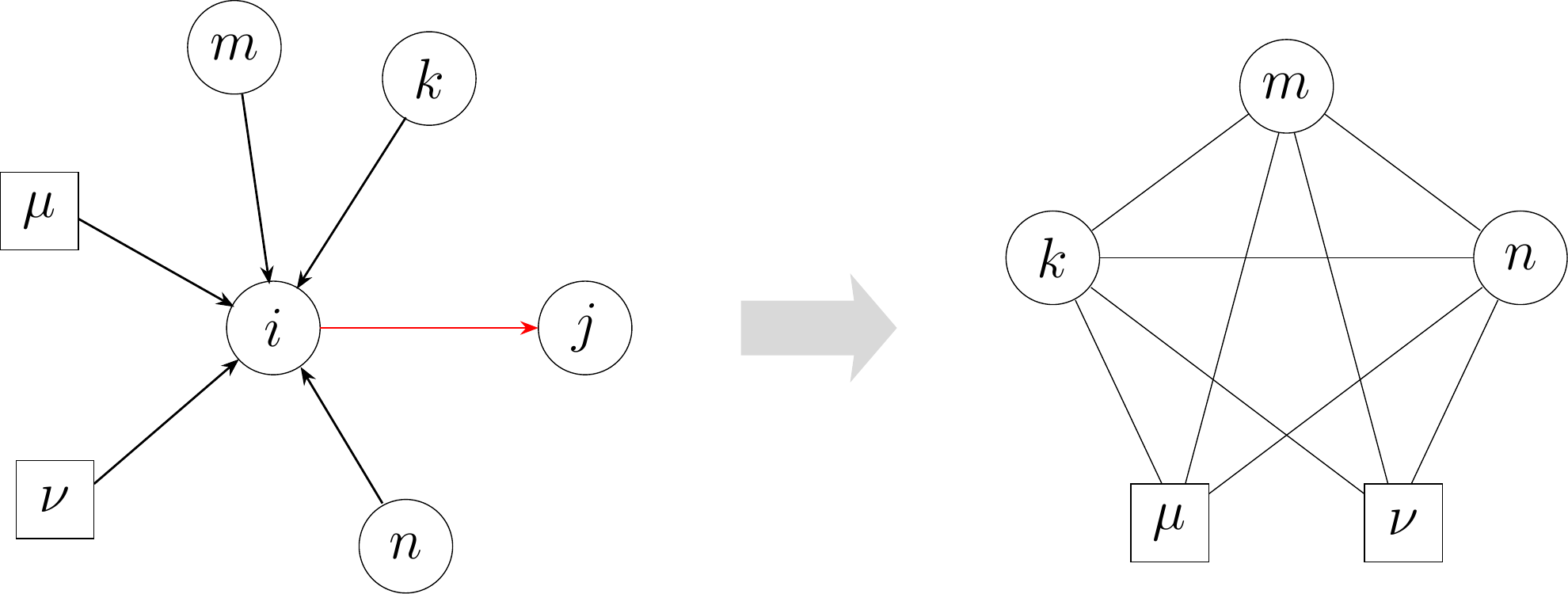}
\caption{\label{fig:ED-MWM}Mapping from EDP  to a maximum weighted matching graph. To calculate the message $\psi_{i\to j}$ by Eq.~\eqref{eq:ED_mess}, we have to consider all valid paired configurations of neighboring nodes \{$k,m,n,\mu,\nu$\}. Transmissions could pass through $m\to i\to n$, so there is an edge ($m,n$) representing the pair interaction, and the contribution to $\phi_{i\to j}$ is approximately $\min_s [\phi_{m\to i}(s)+ \phi_{n\to i}(-s)]$ , whose inverse value is the weight of edge ($m,n$). Auxiliary (square) nodes $\mu$ and $\nu$ represent virtual origins or destinations of different transmissions so there are no paths of the form $\mu\to i\to \nu$, and consequently no edges between them exist.}
\end{figure}

In Fig.~\ref{fig:ED-MWM}, we show the transformation employed for calculating the messages $i\to j$. Links between all pairs, including auxiliary origin-destination and graph edges, are shown on the left. The right figure shows all valid links between nodes, while distinguishing between auxiliary and graph nodes: any of the ordinary graph nodes could be paired but the auxiliary nodes could only be paired with ordinary graph nodes. The maximum matching algorithm finds edge sets with the maximum sum of weights, where matched edges have no common nodes; the inverse value of the weights sum is approximately the value of minimum matched configurations in Eq.~\eqref{eq:ED_mess}.

\subsubsection{Linear cost}

Similar to Sec.~\ref{sec:ND_short}, if the cost function on edges is linear $F_{i,j}(x)=x$, then the message-passing equations could be simplified as follows,
\begin{equation}
\label{eq:ED_mess_short}
\left\{
\begin{aligned}
\phi_{i\to j}^a(0) \sim &~ \min_{\mbox{matched}\atop\mbox{pairs: }\vec s_{\partial i\setminus j}} \sum_{k\in\partial i\setminus j} \phi_{k\to i}^a(s_k),\\
\phi_{i\to j}^a(s) \sim &~ 1+ \min_{k\in\partial i\setminus j} \Big[\phi_{k\to i}^a(s)+ \\
&~ \min_{\mbox{matched}\atop\mbox{pairs: }\vec s_{\partial i\setminus j,k}} \sum_{l\in\partial i\setminus j,k}\phi_{l\to i}^a(s_l)\Big].
\end{aligned}
\right.
\end{equation}
The marginal states of nodes and edges, and messages to the auxiliary nodes at transmission origins and destinations, are the same as in the NDP case of Eq.~\eqref{eq:ND_endpoint} and \eqref{eq:ND_marginal_shortest}.

\subsection{Algorithmic framework\label{sec:algo}}

The same algorithmic procedure governs both NDP and EDP routing algorithms based on the message passing equations. The min-sum algorithm, i.e. zero-temperature optimization algorithm, follows the process outlined below:
\begin{enumerate}
\item initialize the messages $\{\psi_{i\to j}^a, \phi_{i\to j}^a, \phi_{\mu\to a}\}$;
\item update the messages by Eq.~\eqref{eq:ND_mess}--\eqref{eq:ND_endpoint} in the node-disjoint scenarios, and Eq.~\eqref{eq:ED_mess} and \eqref{eq:ND_endpoint} in the edge-disjoint scenarios until convergence or when a maximum number of iteration steps is reached;
\item calculate the marginal state of edges by Eq.~\eqref{eq:ND_marginal} and the state of each wavelength channel by Eq.~\eqref{eq:ND_state_chooser}.
\end{enumerate}
After obtaining the marginal states of edges and wavelengths, the allocation of transmissions to specific wavelength and path would follow.

However, there are some cases when the algorithms need many iterations to provide a valid configuration, for example if the cost function $F_{i,j}(x)$ in Eq.~\eqref{eq:hamilton} is concave, e.g. $F(x)= \sqrt{x}$ in these cases, a decimation procedure can be introduced to speed up convergence. As an example, here we show the NDP algorithm with decimation (fixing states at intermediate steps) incorporated as part of the process:
\begin{enumerate}
\item initialize the messages $\{\psi_{i\to j}^a, \phi_{i\to j}^a, \phi_{\mu\to a}\}$ and the state of all wavelength channels per available (undecimated) edge;
\item \label{step:deci1}update the messages of the available wavelength channels and edges by Eq.~\eqref{eq:ND_mess}--\eqref{eq:ND_endpoint} for a specific number of iteration steps;
\item \label{step:deci2}compute the marginal states of the available wavelength channels and edges by Eq.~\eqref{eq:ND_marginal}; calculate the quantity $\hat\phi_{i,j}^a= \min_{s\neq 0}\phi_{i,j}^a(s)- \phi_{i,j}^a(0)$, and fix the state of the wavelength channel $a$ of edge $(i,j)$ with the largest value of $\hat\phi_{i,j}^a$  to be $s_{i,j}^a=0$.
\item iterate step~\ref{step:deci1} and \ref{step:deci2} and determine the state of the available  wavelength channels and edges by Eq.~\eqref{eq:ND_state_chooser} until a valid solution is obtained or a maximum number of iteration steps is reached.
\end{enumerate}

The algorithms with or without the decimation procedures have a similar performance in terms of the optimized cost, but for harder problem (e.g. $F(x)= \sqrt{x}$) the decimation procedures can reduce the number of iteration steps needed for generating a valid routing solution.

\section{Simulation results\label{sec:results}}

\subsection{Linear cost}

We first examine  the NDP, {WS} and EDP scenarios with linear cost $F(x)=x$ to explore the behavior and performance of communication networks, such as capacity and average length, with impact on latency and number of wavelength channels required. 

\subsubsection{Dependence on the number of wavelength channels\label{sec:diff_wave}}

\begin{figure}[hbtp!]
\centering
\subfigure[CONUS]{\includegraphics[height=0.165\textwidth]{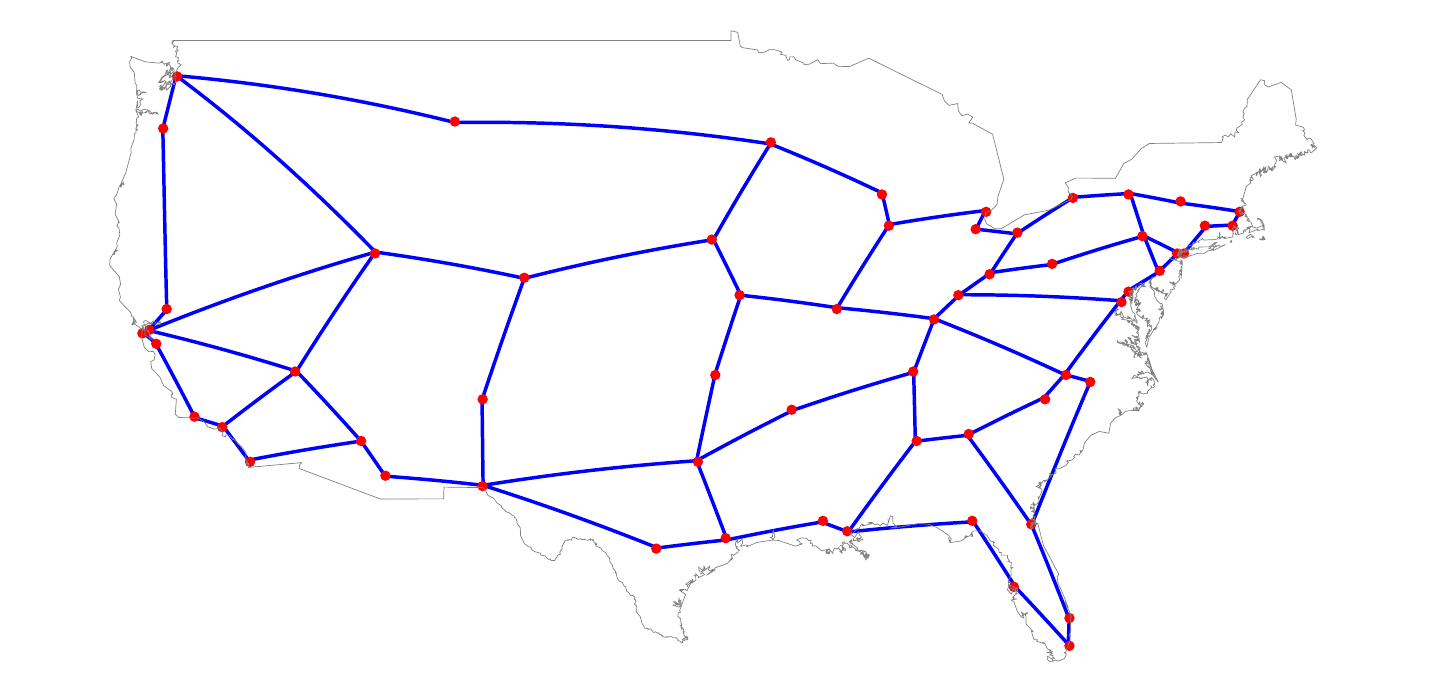}}
\subfigure[BT-Core]{\includegraphics[height=0.165\textwidth]{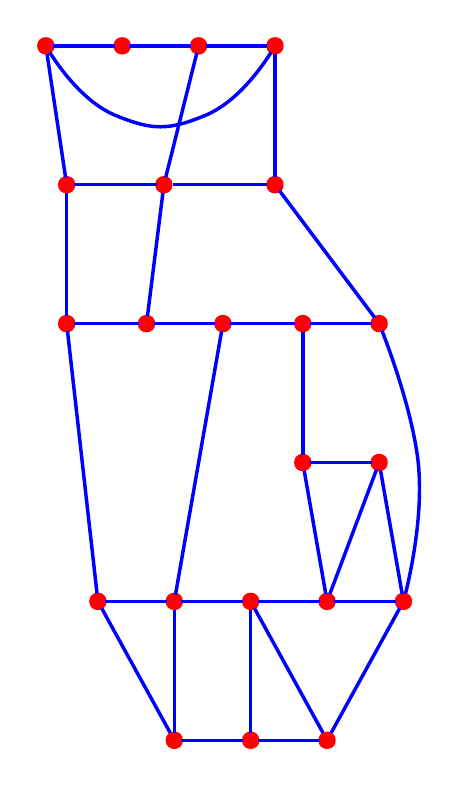}}
\caption{\label{fig:mapNet}Real optical communication networks studied, namely (a) CONUS60 in the United States, and (b) BT-core in the United Kingdom.}
\end{figure}

We performed numerical experiments using the three algorithms on three different types of generated random networks including random regular graphs, \ER \ and scale-free networks with $100$ nodes and an average node degree of $3$, as well as on two real optical communication networks --- CONUS60, which has $60$ nodes and $79$ edges~\cite{conus60} and BT-Core, $22$ nodes and $35$ edges~\cite{wright2012} as shown in Fig.~\ref{fig:mapNet}. Specifically, for the scale-free networks studied, the degree distribution is $p(d)\sim d^{-2.63}$.

\begin{figure*}[hbt!]
\centering
\subfigure[Node-disjoint]{\includegraphics[width=0.32\textwidth]{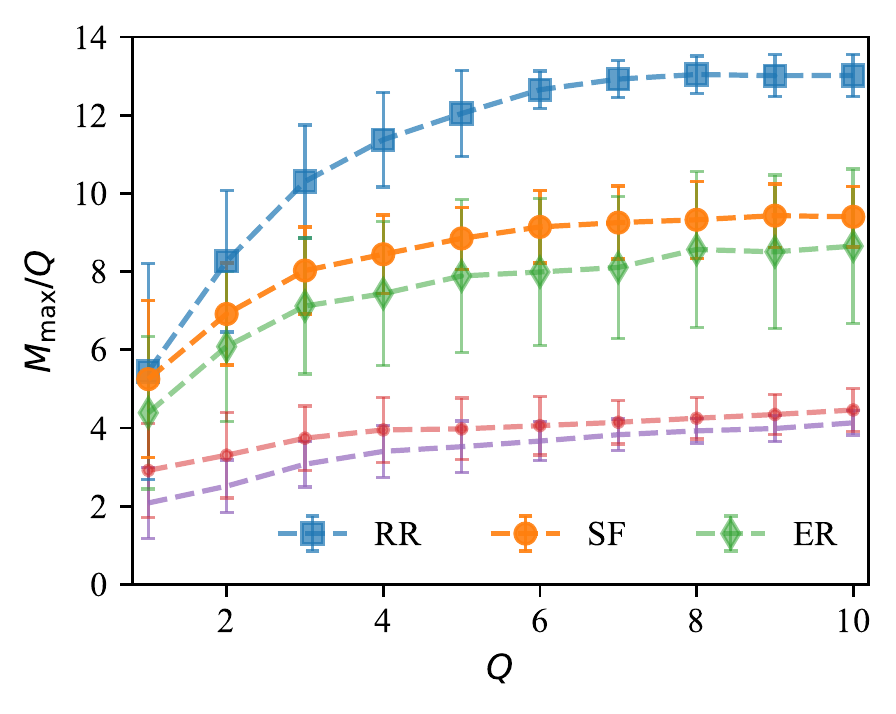}
\label{fig:nodeDisjointCapacity}}
\subfigure[Node-disjoint wavelength-switching]{\includegraphics[width=0.32\textwidth]{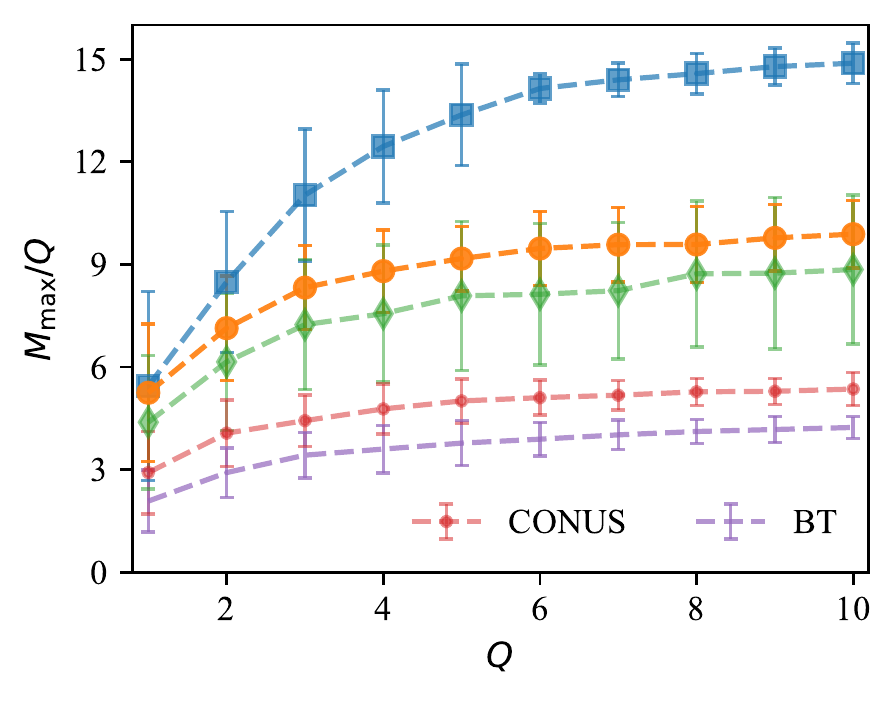}
\label{fig:nodeSwitchingCapacity}}
\subfigure[Edge-disjoint]{\includegraphics[width=0.32\textwidth]{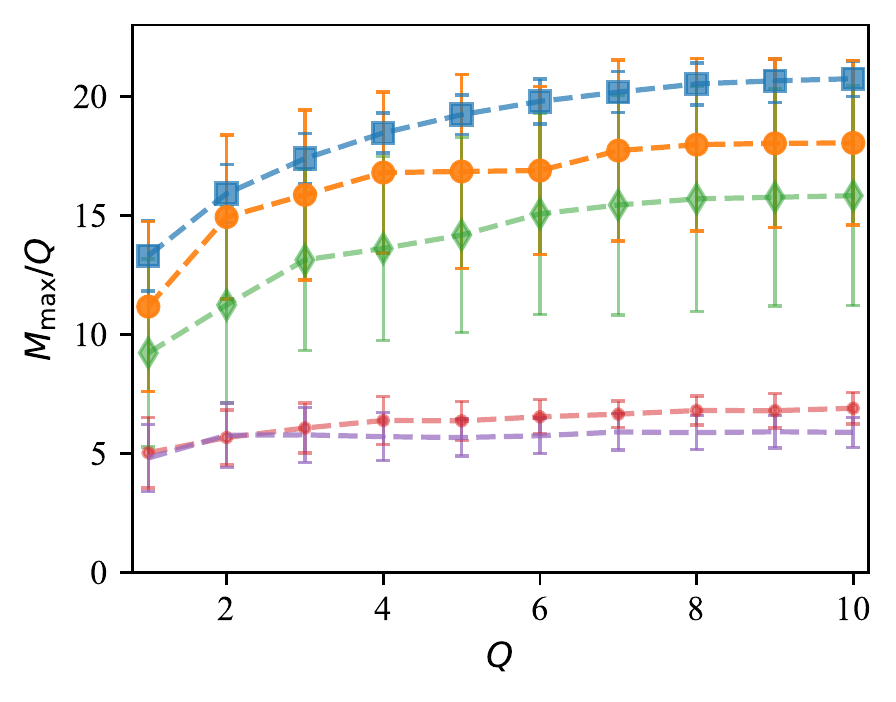}
\label{fig:edgeDisjointCapacity}}
\caption{\label{fig:Capacity}The capacity per wavelength channel, i.e. $M_{\rm max}/Q$, as a function of $Q$, in the NDP, wavelength-switching and EDP scenarios with linear cost, on random regular (RR), \ER \ (ER) and scale-free (SF) networks, as well as a real optical communication networks known as CONUS and BT~\cite{conus60,wright2012}. All three types of generated networks have $100$ nodes and an average degree of $3$, while the CONUS network has $60$ node and $79$ edges, and BT $22$ nodes and $35$ edges. All the results are obtained by averaging $36$ realizations.}
\end{figure*}

We define the \emph{capacity} to be the maximum number of transmissions, which we denote as $M_{\rm max}$, that can be transmitted by an optical communication network with $Q$ wavelength channels. We remark that capacity depends on network topologies and the set of origin-destination pairs, so in Fig.~\ref{fig:Capacity} we report the dependence of the average value of $M_{\rm max}$ with standard deviation as error-bars on $Q$, showing the average behavior of capacities for different $Q$ values.

Fig.~\ref{fig:Capacity} shows the simulation results for the five types of networks --- in the three different routing scenarios. For the three types of random networks and BT-Core, the average capacity per wavelength channel $M_{\rm max}/Q$ keeps increasing at the beginning and become saturated as the number of wavelength channels $Q$ increases.As for the networks CONUS and BT-Core in Fig.~\ref{fig:Capacity}, we see that the average capacities also increase with $Q$ but the increases are not as fast as that in the three generated networks.

\begin{figure}[hbt!]
\centering
\subfigure[Random regular]{
\includegraphics[width=0.226\textwidth]{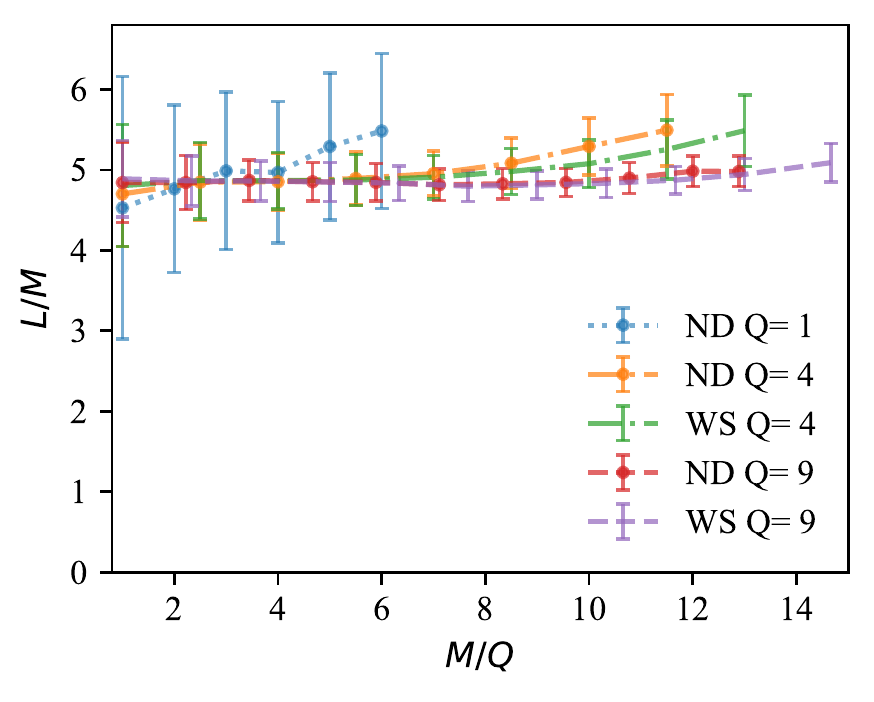}
\label{fig:NDLengthRR}}
\subfigure[\ER]{
\includegraphics[width=0.226\textwidth]{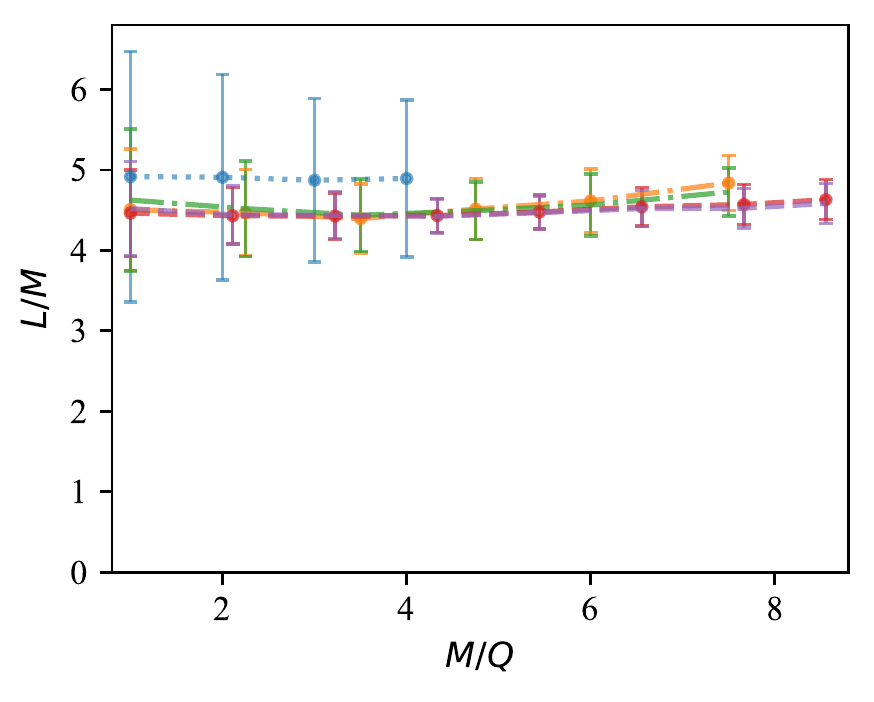}
\label{fig:NDLengthER}}
\subfigure[Scale-free]{
\includegraphics[width=0.226\textwidth]{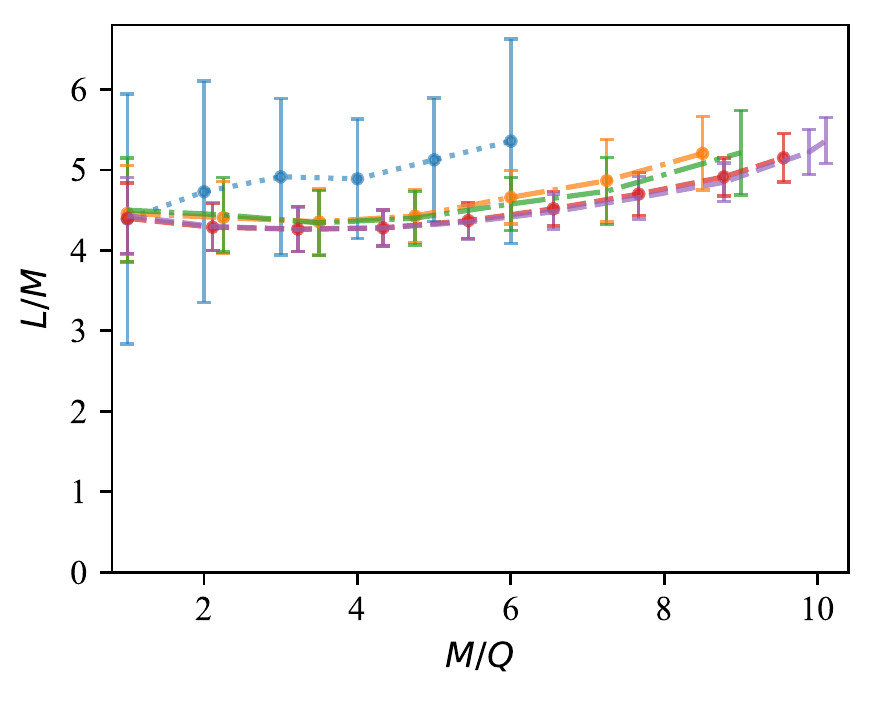}
\label{fig:NDLengthSF}}
\subfigure[CONUS]{
\includegraphics[width=0.226\textwidth]{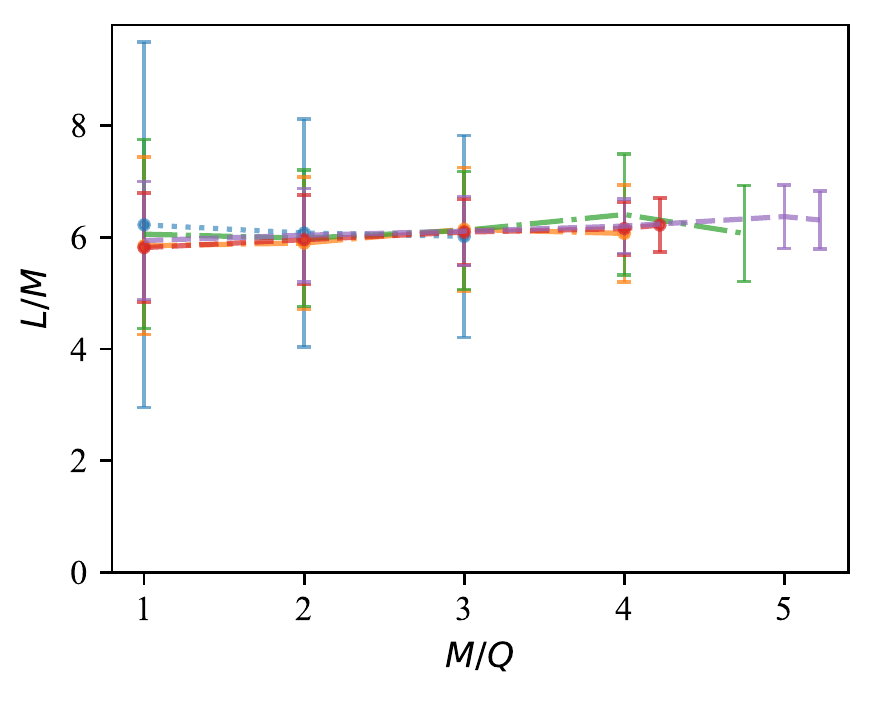}
\label{fig:NDLengthUS}}
\caption{\label{fig:ND-Cap-len}The dependence of the path length $L/M$ on the number of transmissions per wavelength $M/Q$ for different $Q$ values, in the NDP and wavelength-switching (WS) scenarios with a linear cost, and for the four types of graphs studied. The results are obtained by averaging no less than $20$ samples. The WS scenario does not show a significant advantage over the original node-disjoint scenario for \ER\ networks based on the limited simulation results.}
\end{figure}

\begin{figure}[hbt!]
\centering
\subfigure[Random regular]{
\includegraphics[width=0.226\textwidth]{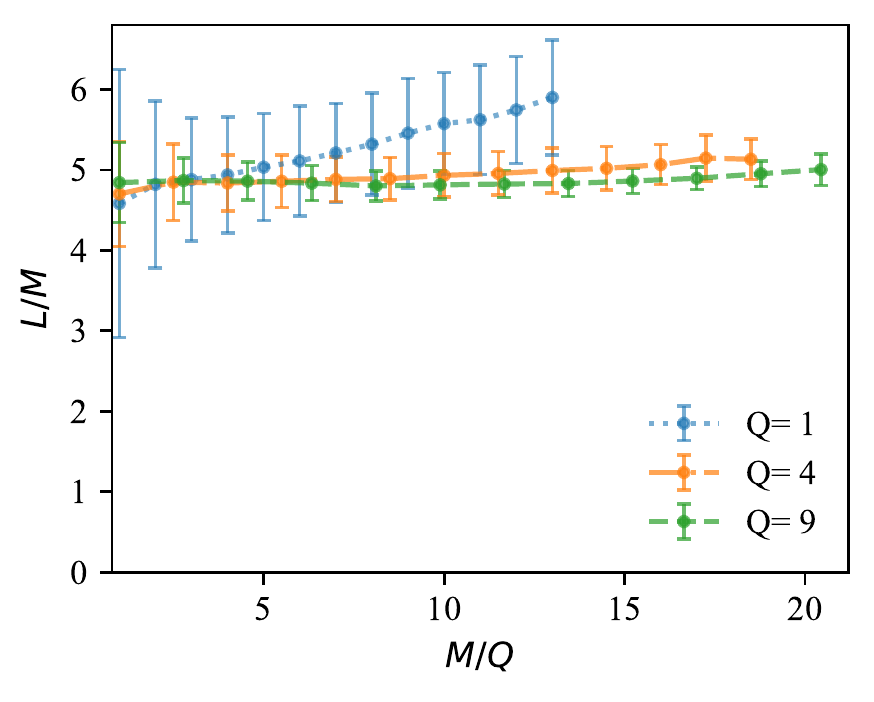}
\label{fig:EDLengthRR}}
\subfigure[\ER]{
\includegraphics[width=0.226\textwidth]{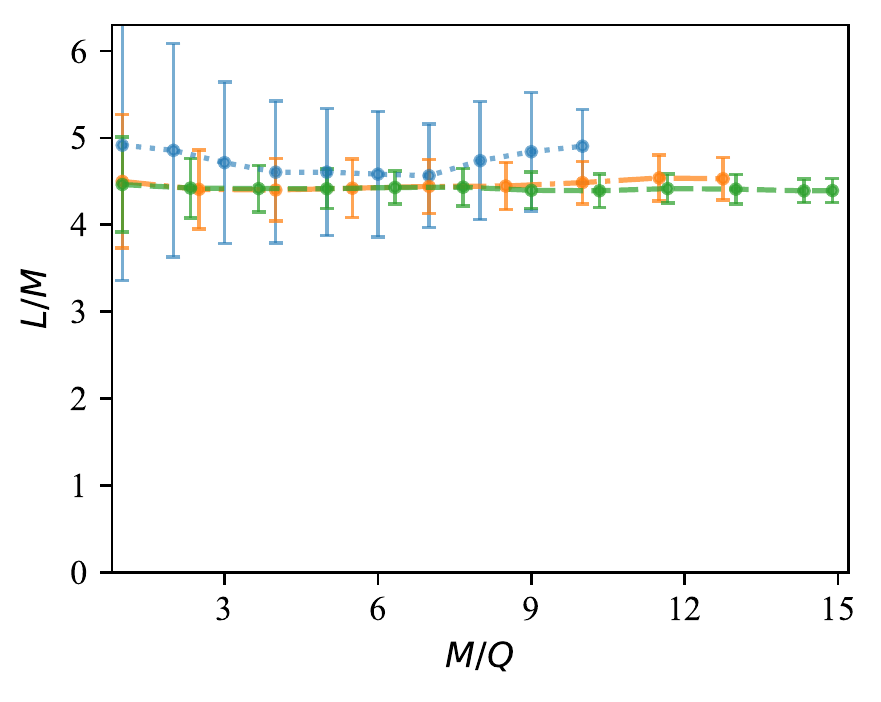}
\label{fig:EDLengthER}}
\subfigure[Scale-free]{
\includegraphics[width=0.226\textwidth]{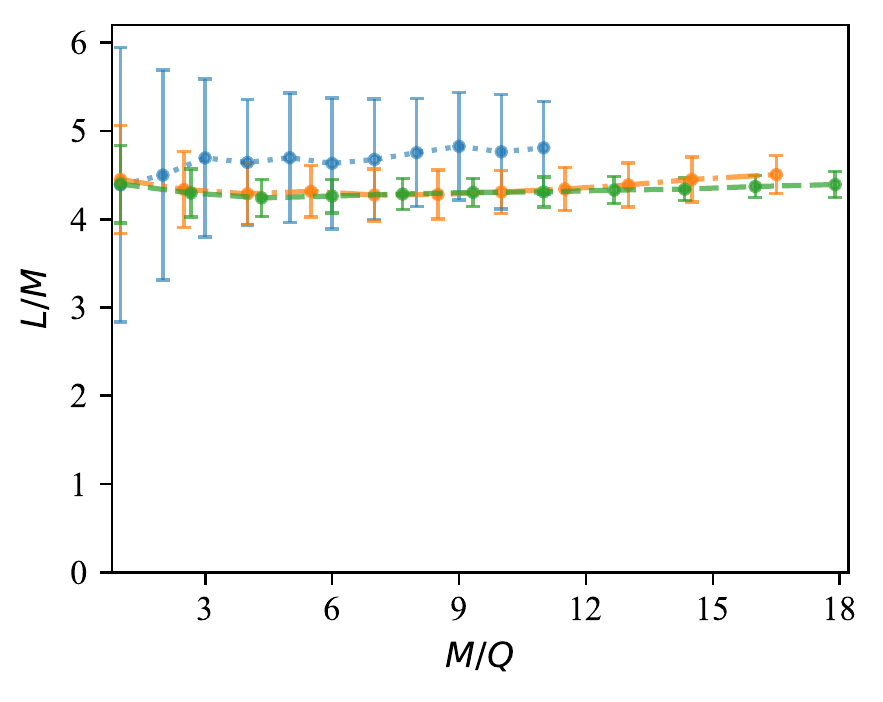}
\label{fig:EDLengthSF}}
\subfigure[CONUS]{
\includegraphics[width=0.226\textwidth]{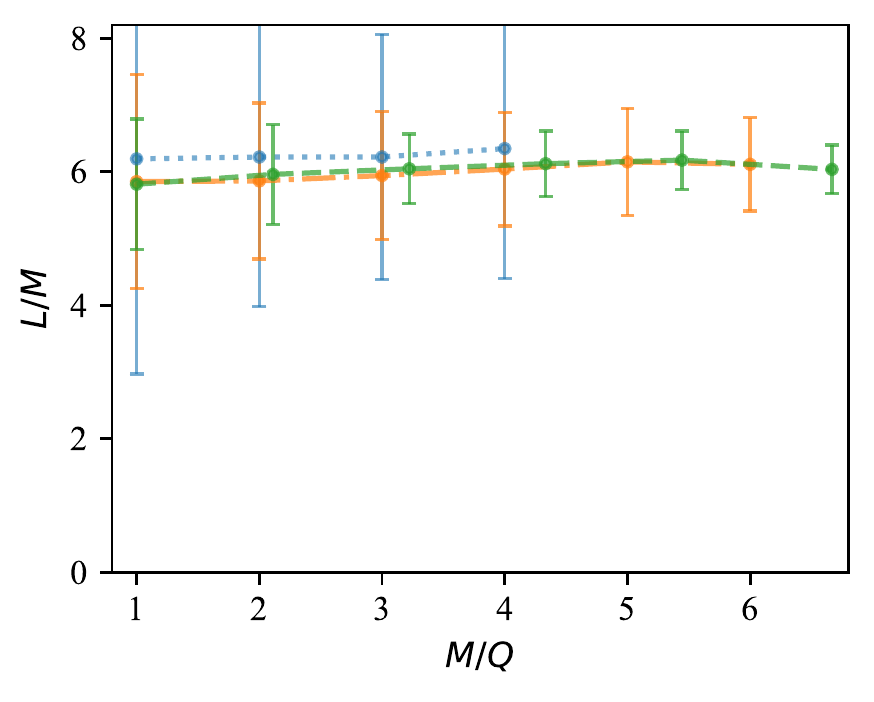}
\label{fig:EDLengthUS}}
\caption{\label{fig:ED-Cap-len}The dependence of the path length $L/M$ on the number of transmissions per wavelength $M/Q$ for different values of $Q$, in EDP scenarios with a linear cost, and for the four types of graphs studied. The results are obtained by averaging no less than $20$ samples per point.}
\end{figure}

The average path length of the corresponding networks are shown in Fig.~\ref{fig:ND-Cap-len} and \ref{fig:ED-Cap-len}. In Fig.~\ref{fig:ND-Cap-len}, we see that as more wavelengths are available the shorter the average path length $L/M$ decreases; for instance, in NDP scenarios shown in Fig.~\ref{fig:NDLengthRR}--\ref{fig:NDLengthSF}, the average path length with $Q=1$ is higher than that with $Q=4$, which is higher than that of $Q=9$ as the average load per wavelength channel ($M/Q$) increases. Moreover, with the same values of $Q$ and $M$, it is clear that the wavelength-switching model would provide us with shorter paths and a higher capacity {on \ER \ and scale-free networks} as shown in Fig.~\ref{fig:ND-Cap-len}. 

In comparison with NDP, the advantages of EDP with  multiple transmissions using the same wavelength availability is clearer on random regular graphs as shown in Fig.~\ref{fig:EDLengthRR}, and less so on other graph types, presumably due to the graph heterogeneity and finite size effects. We observe that EDP routing increases the network capacity, i.e. average transmission load per wavelength channel, and provides shorter valid paths.

\subsubsection{Dependence on the number of transmissions}

In practice, one may wish to minimize the total number of active wavelength channels for given demand. We define the smallest number of wavelength channels which accommodate a set of transmission to be $Q_{\min}$; to find $Q_{\min}$ from for a specific instance, we gradually increase $Q$ from $1$ until a valid solution is found by the algorithms. Here we compare the results of $Q_{\min}$ obtained by our proposed multi-wavelength routing {(MWR)} algorithms to those obtained by a multi-trial greedy assignment (MGA) algorithm in the node-disjoint or edge-disjoint scenarios following the process outlined below:

\begin{enumerate}
\item initialize the values of $\Delta M$, $M$, $Q$ and $M_*$, where $M$ to $M_*$ represent the range of transmission number values we want to explore and $\Delta M$ the step size increase in the experiment;
\item \label{step:RA_2}read the first $M$ elements of the random origin-destination pair list, set $Q= \max(1,~Q- 4)$ to increase the probability of finding a solution in the case of small $Q$;
\item \label{step:RA_3}randomly assign the $M$ transmissions into $Q$ wavelength channels, and solve the routing problem for each transmission on the individually assigned wavelength channel separately;
\item \label{step:RA_4}repeat step~\ref{step:RA_3} up to a maximum number of trials (e.g. $10$)  until a valid configuration is found;
\item \label{step:RA_5}if step~\ref{step:RA_4} fails, set $Q:= Q+ 1$ and repeat until a valid configuration is obtained, then the value of $Q$ is the smallest number of wavelengths $Q_{\min}$ that accommodates the $M$ transmissions;
\item increase $M:= M+ \Delta M$ and repeat step~\ref{step:RA_2}~--~\ref{step:RA_5} until $M\geq M_*$. 
\end{enumerate}

\begin{figure}[hbt!]
\centering
\subfigure[\label{fig:RR_RMS_reslut_NDQ}]{\includegraphics[width=0.226\textwidth]{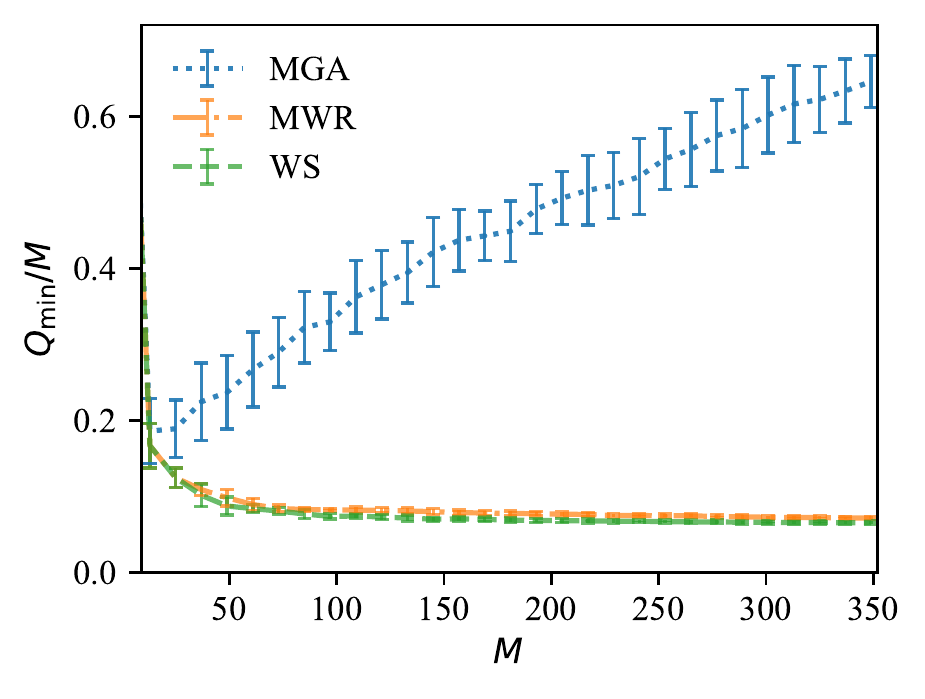}}
\subfigure[\label{fig:RR_RMS_reslut_NDL}]{\includegraphics[width=0.222\textwidth]{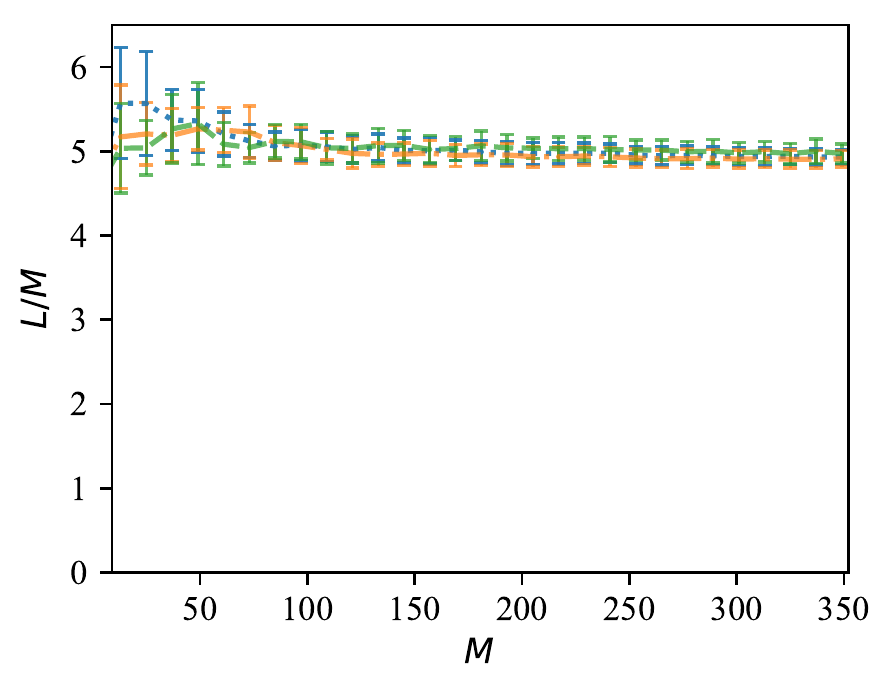}}
\subfigure[\label{fig:RR_RMS_reslut_EDQ}]{\includegraphics[width=0.226\textwidth]{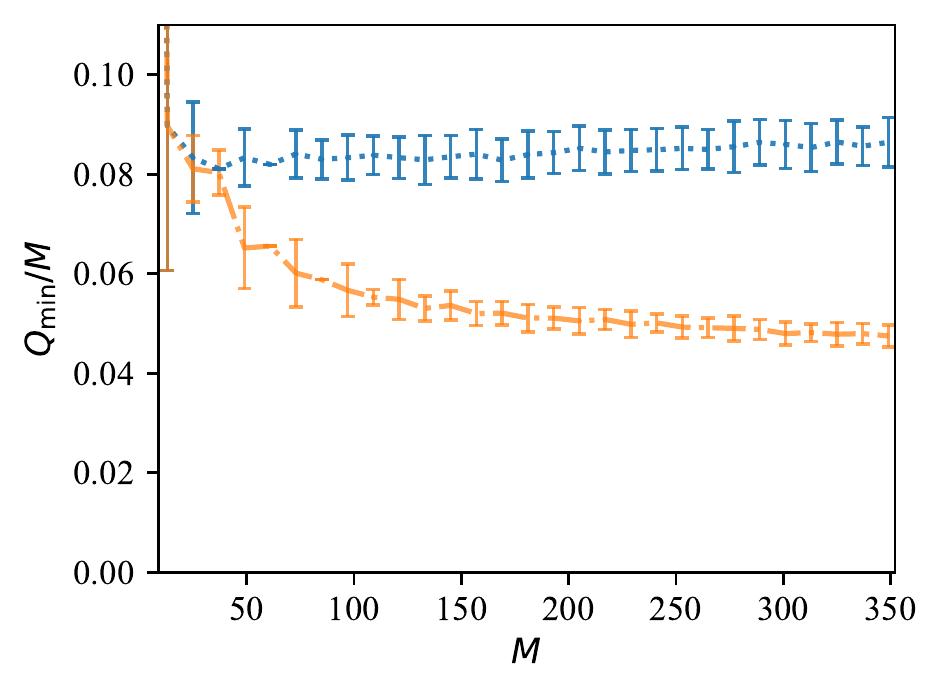}}
\subfigure[\label{fig:RR_RMS_reslut_EDL}]{\includegraphics[width=0.22\textwidth]{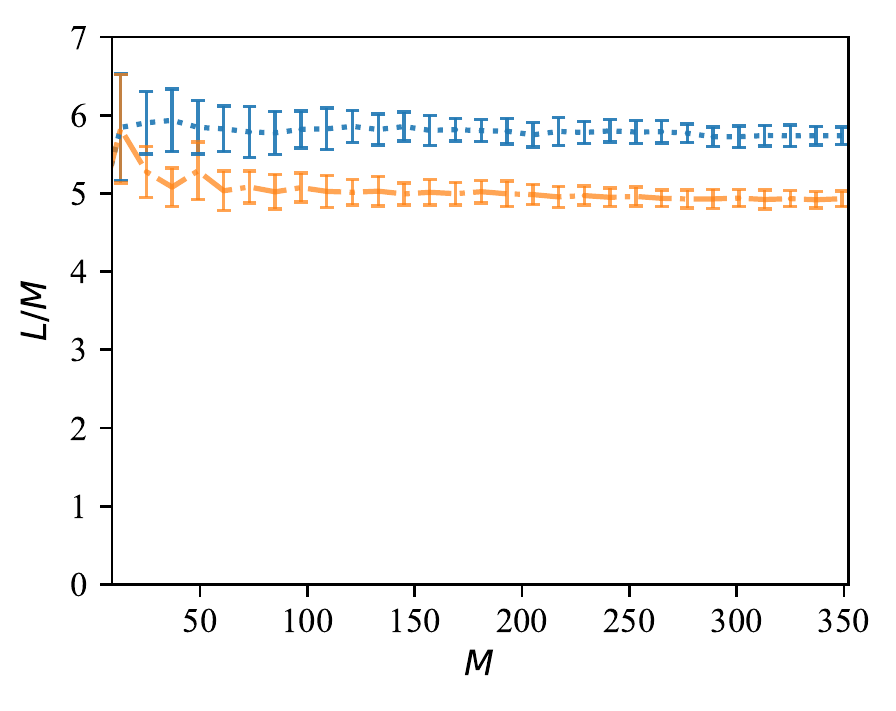}}
\caption{\label{fig:RR_RMS_reslut} The average smallest number of wavelength channels $Q_{\min}/M$ with (a) NDP and (c) EDP routing, and the average path length with (b) NDP and (d) EDP routing, as a function of the number of transmissions $M$, on random regular networks with $100$ nodes and degree $3$, obtained by our multi-wavelength routing algorithm with NDP{/EDP (MWR,} orange) and WS (green), in comparison with the multi-trial greedy assignment algorithm (MGA, blue). The results are obtained by averaging $36$ samples.}
\end{figure}

\begin{figure}[hbtp!]
\centering
\subfigure[]{\includegraphics[width=0.226\textwidth]{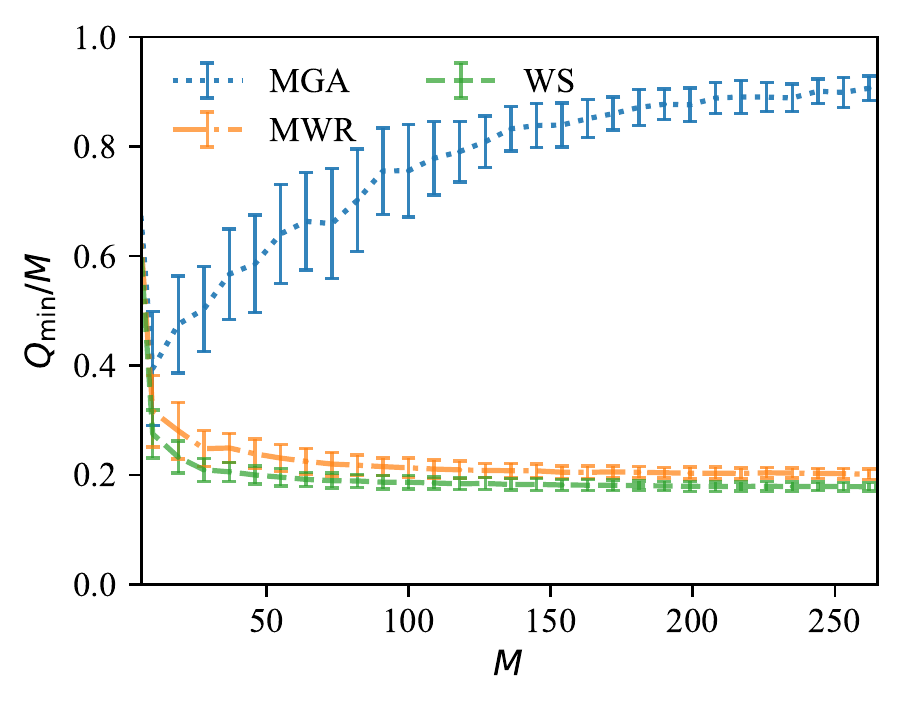}}
\subfigure[]{\includegraphics[width=0.222\textwidth]{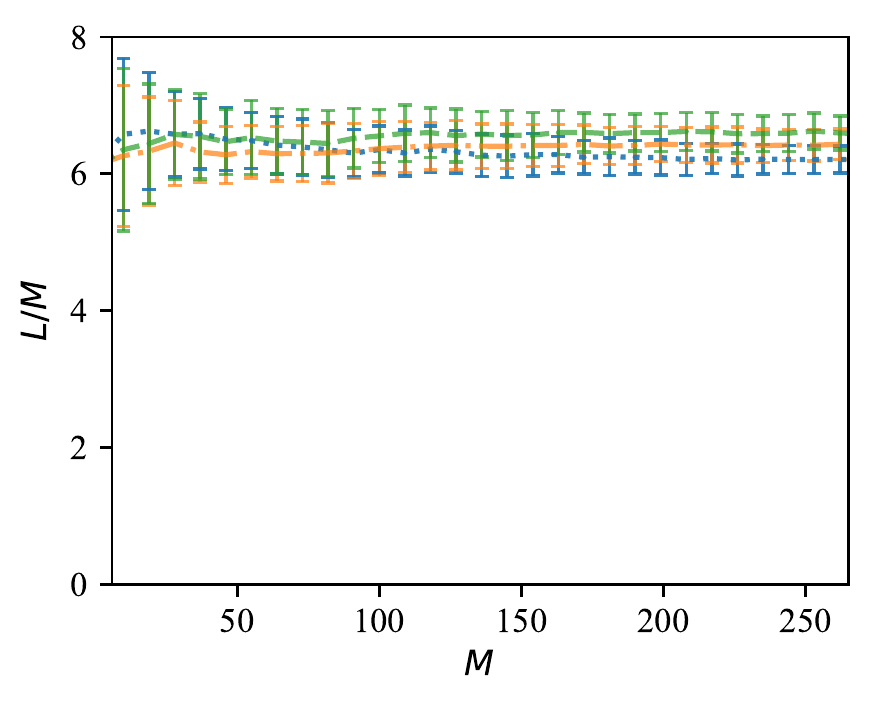}}
\subfigure[]{\includegraphics[width=0.226\textwidth]{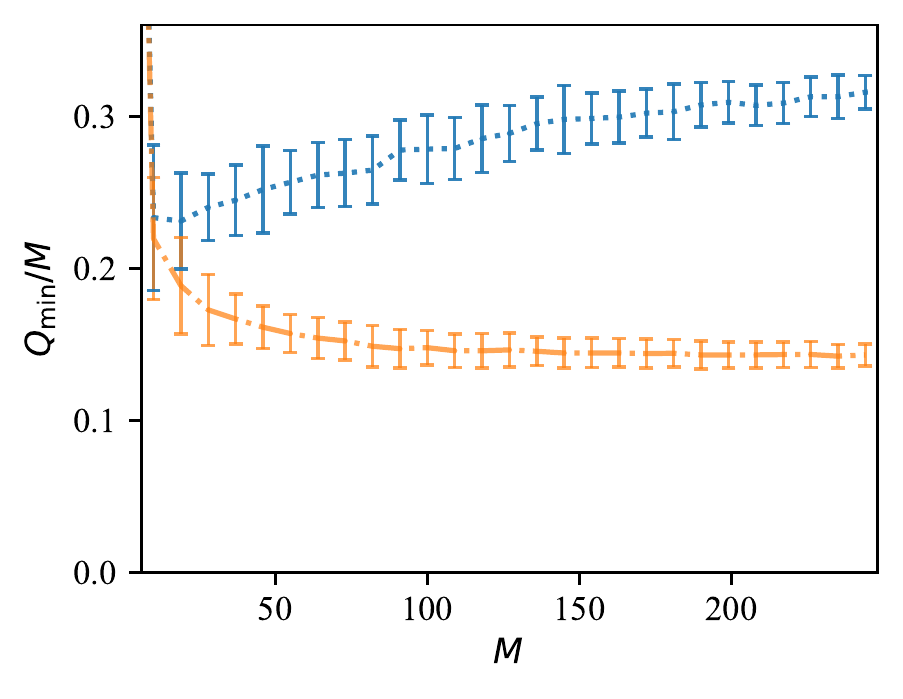}}
\subfigure[]{\includegraphics[width=0.22\textwidth]{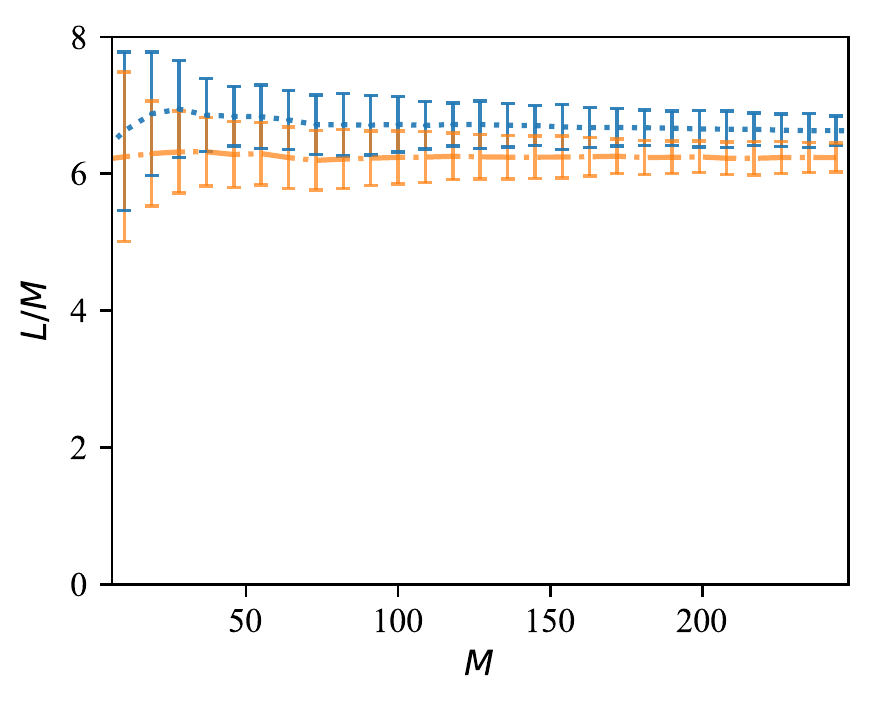}}
\caption{\label{fig:CONUS_RMS_reslut}The average smallest number of wavelength channels $Q_{\min}/M$ with (a) NDP and (c) EDP routing, and the total path length with (b) NDP and (d) EDP routing, as a function of the number of transmissions $M$, on the CONUS network, obtained by our MWR algorithm with NDP/EDP (orange) and WS (green), in comparison with MGA algorithm (blue). The results are obtained by averaging $36$ samples.}
\end{figure}

\begin{figure}[hbtp!]
\centering
\subfigure[]{\includegraphics[width=0.226\textwidth]{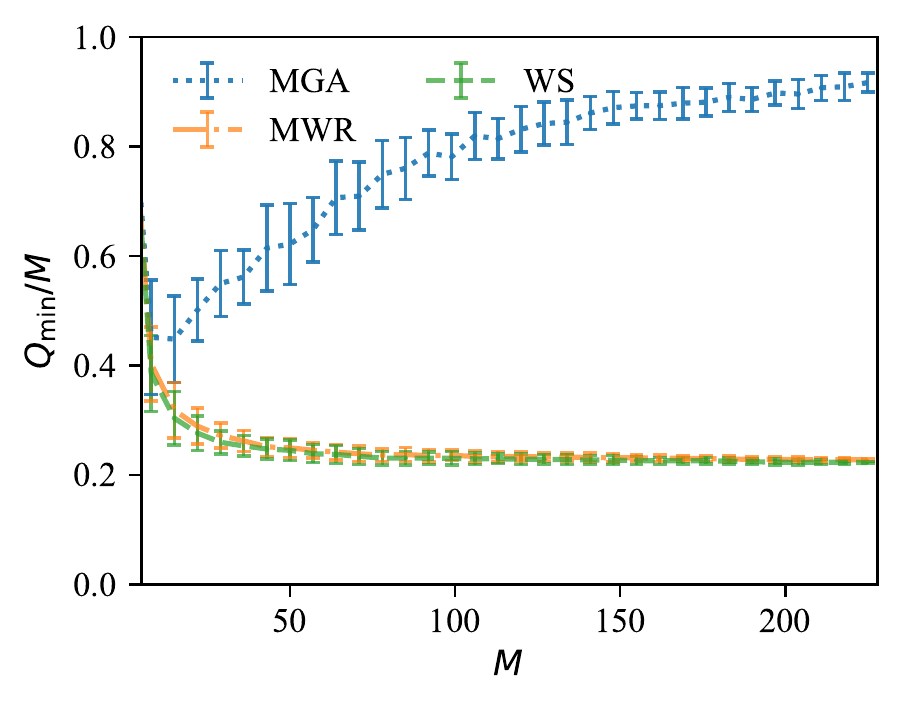}}
\subfigure[]{\includegraphics[width=0.222\textwidth]{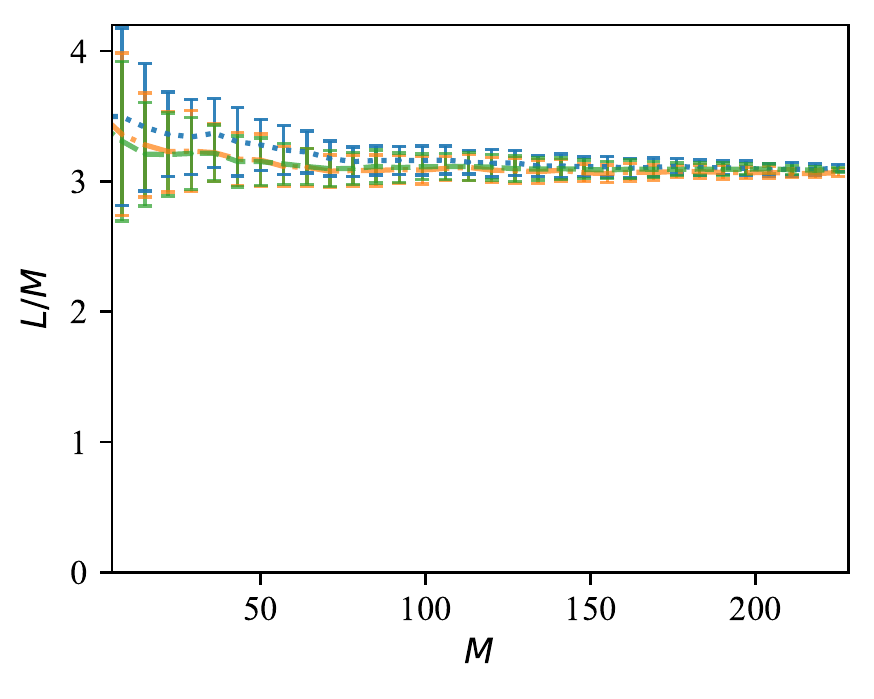}}
\subfigure[]{\includegraphics[width=0.226\textwidth]{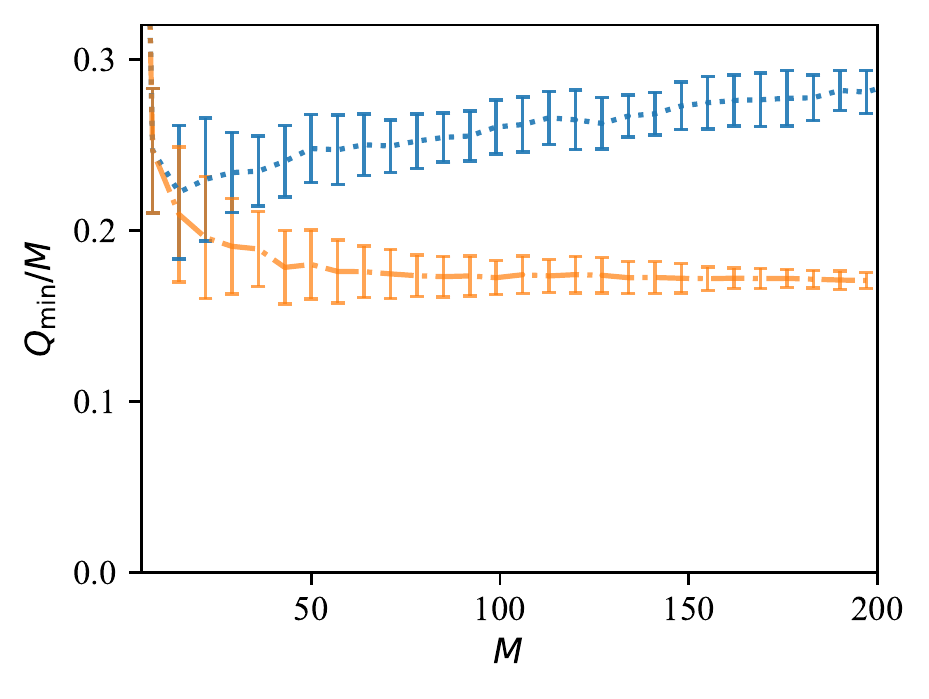}}
\subfigure[]{\includegraphics[width=0.22\textwidth]{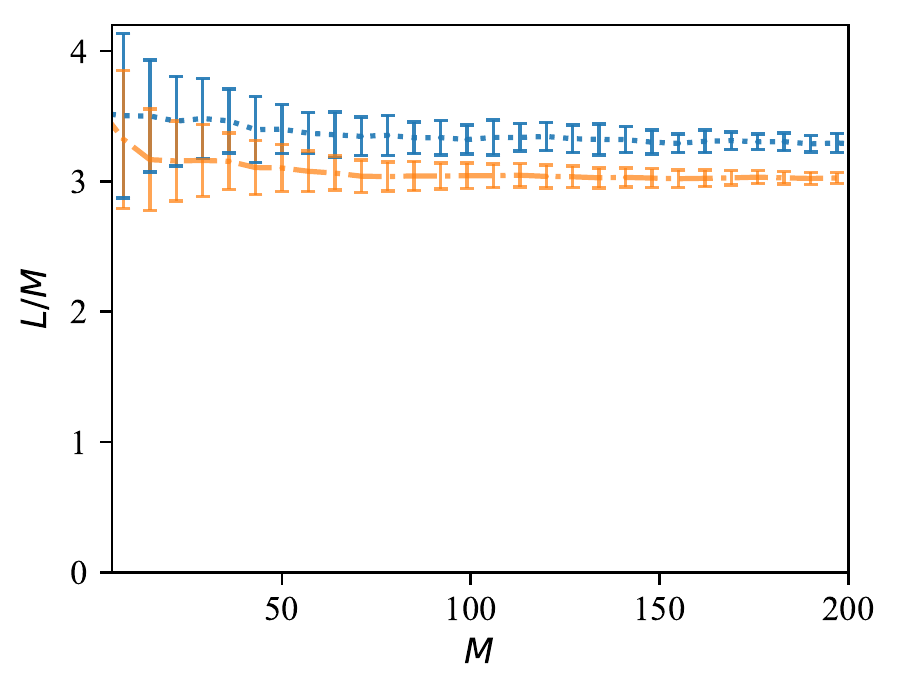}}
\caption{\label{fig:BT_RMS_reslut}The average smallest number of wavelength channels $Q_{\min}/M$ with (a) NDP and (c) EDP routing, and the total path length with (b) NDP and (d) EDP routing, as a function of the number of transmissions $M$, on the BT-core network, obtained by our MWR algorithm with NDP/EDP (orange) and WS (green), in comparison with MGA algorithm (blue). The results are obtained by averaging $36$ samples.}
\end{figure}

We compare the values of $Q_{\min}/M$ obtained by our algorithms and by the MGA algorithm on random regular networks, the CONUS and BT-Core networks, in both NDP and EDP scenarios. The numerical results on random regular networks presented in Fig.~\ref{fig:RR_RMS_reslut_NDQ} and \ref{fig:RR_RMS_reslut_EDQ} show that our algorithms offer significant advantages over the MGA algorithm in reducing the number of wavelength channels required for specific random transmission pairs in. The improved performance is also observed on the CONUS and BT-Core networks in Fig.~\ref{fig:CONUS_RMS_reslut} and \ref{fig:BT_RMS_reslut} respectively. The underlying reason for the improvement is the low success rate of random greedy assignments. The numerical results demonstrate that our MWR algorithms lead to a better use of resource. The additional flexibility provided by the wavelength-switching on transceivers leads to an even smaller value of $Q_{\min}$ as $M$ increases in Fig.~\ref{fig:RR_RMS_reslut_NDQ} and \ref{fig:CONUS_RMS_reslut}(a), whereas the experiments on the BT-Core network do not show {a significant improvement of WS over the ordinary NDP} in Fig.~\ref{fig:BT_RMS_reslut}(a), which may depend on the topology of the graph or our range of values tested.

Comparing the average path length found for both NDP and EDP routing on random regular graphs, we see in Fig.~\ref{fig:RR_RMS_reslut_NDL} that the average path lengths obtained by the MGA and MWR for NDP routing are similar but that shorter paths are found by MWR in the EDP scenario (Fig.~\ref{fig:RR_RMS_reslut_EDL}). A significant reduction in average route length is not expected since typical route lengths on random graphss is $O(\log N)$ with little variability. However, a significant reduction in the number of wavelength channels used $Q_{\min}$, is shown in Fig.~\ref{fig:RR_RMS_reslut_EDL} and a similar trend is shown for EDP scenarios in Fig.~\ref{fig:CONUS_RMS_reslut}(a,c) and \ref{fig:BT_RMS_reslut}(a,c) for the two real networks.

\subsection{Computational complexity\label{sec:complexity}}

Next, we discuss the computational complexity of our multi-wavelength routing algorithms with a linear cost function $F_{i,j}(x)=x$. In this case, if we use the normalization $\phi^a_{i\to j}(s):= \phi^a_{i\to j}(s)- \phi^a_{i\to j}(0)$ in the  min-sum equations~\eqref{eq:ND_mess_shortest}\eqref{eq:NS_mess}\eqref{eq:ED_mess}, the computational complexities would decrease.

\begin{figure*}[hbt!]
\centering
\subfigure[Node-disjoint]{
\includegraphics[width=0.30\textwidth]{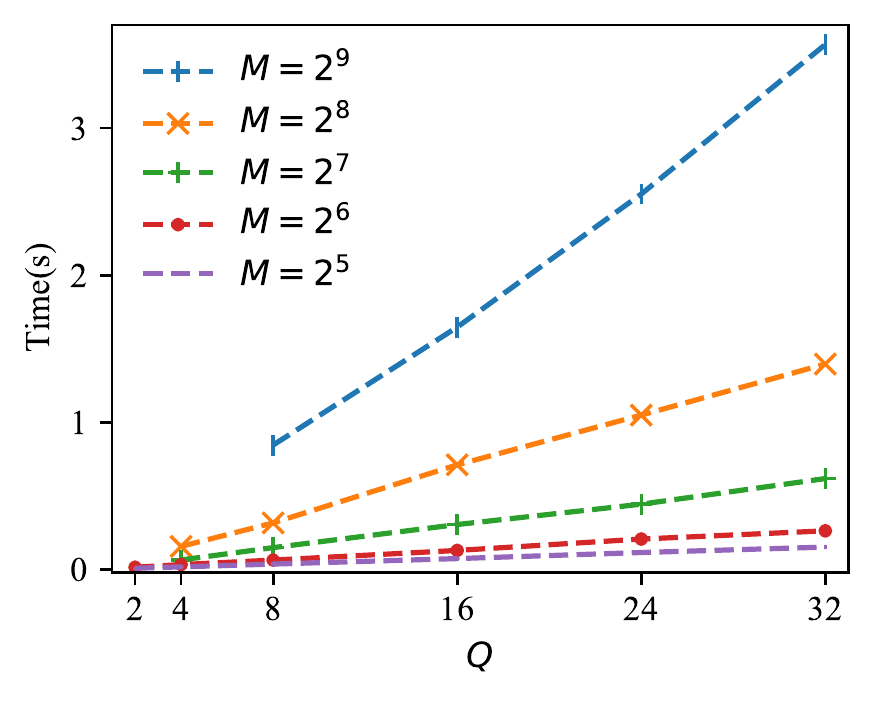}
\label{fig:NDrrComplex}}
\subfigure[Edge-disjoint]{
\includegraphics[width=0.31\textwidth]{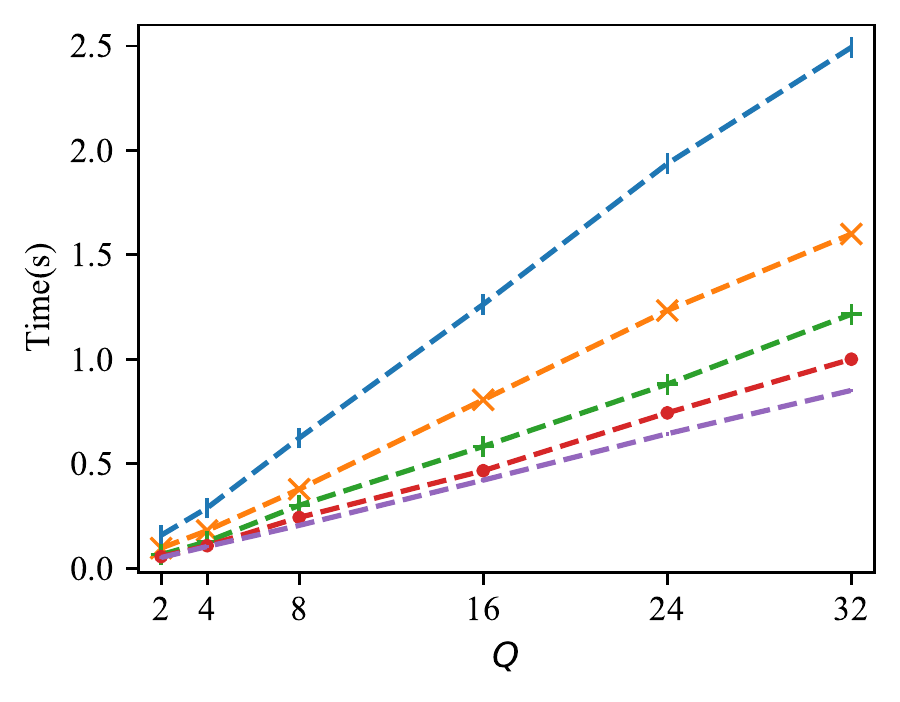}
\label{fig:EDrrComplex}}
\subfigure[Large $M$ complexity]{
\includegraphics[width=0.32\textwidth]{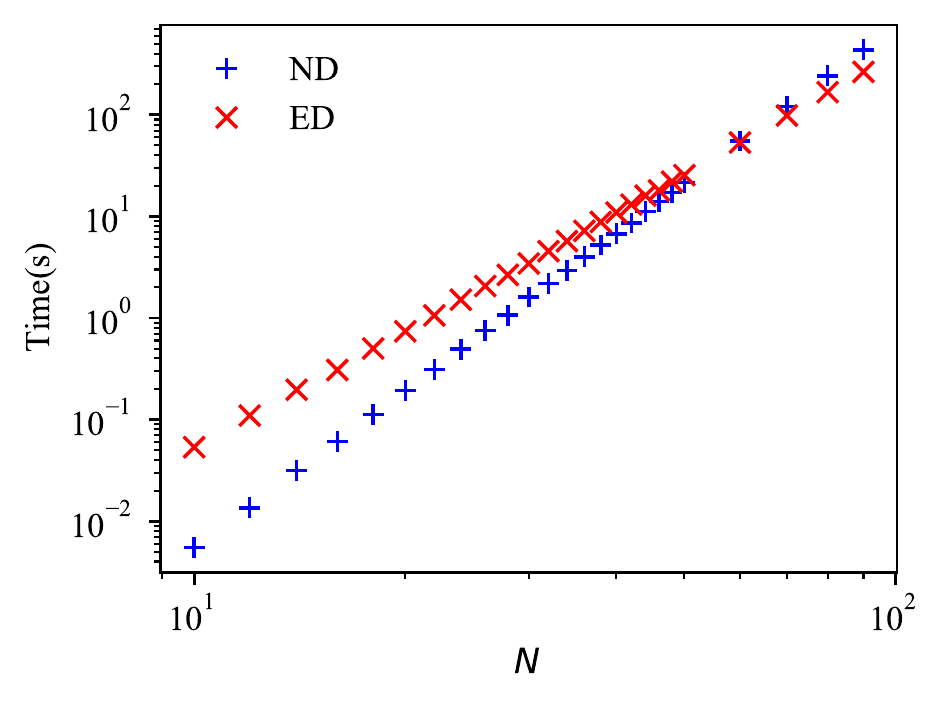}
\label{fig:NComplex}}
\caption{\label{fig:complexity} The dependence of one round of computational time costs from our multi-wavelength routing algorithms on the number of wavelengths $Q$. For relatively small $Q\ll N$ and $M\ll N^2$, data obtained on random regular networks with $1000$ nodes and degree $3$ in (a) NDP and (b) EDP scenarios. With the same number of transmissions $M$, the time costs scale roughly linearly with $Q$ for both scenarios. For large $M$, when allocate all the $M= N(N-1)/2$ transmissions on $Q= N$ wavelengths, the log-log plots of one round computation time and node size $N$ are presented in (c). It is easy to draw straight lines to fit the data points and the fitting slops are around $3.92$ and $5.14$ for EDP and NDP scenarios respectively. The simulations are implemented on random regular networks of degree $3$ and node size $N$.}
\end{figure*}

Under NDP routing, the complexity in computing Eq.~\eqref{eq:ND_mess_shortest} is approximately $O(\langle k\rangle^2 M)$, where $\langle k\rangle$ is the average node degree, and there are $\langle k \rangle N Q$ such messages. The complexity in computing the messages from origin/destination nodes in \eqref{eq:ND_endpoint} is $O(Q)$, and there are $2MQ$ such messages. Therefore, the total complexity of one round of update for all messages is $O(\langle k \rangle^3 MNQ+ MQ^2)$; if the graph is sparse, i.e. $\langle k\rangle \ll N$, the complexity becomes $O(MNQ+ MQ^2)$. For large graphs and relatively small $Q\ll N$, the complexity is roughly $O(MNQ)$, which linearly scales with $Q$ as shown in Fig.~\ref{fig:NDrrComplex}.

For EDP routing, the complexity of matching a pair of incoming and outgoing transmissions with the same wavelength in Eq.\eqref{eq:ED_mess_short} is $O(\langle k\rangle^3 \log \langle k\rangle)$ ~\cite{altarelli2015edge}. Therefore, the total complexity of one round of update for all messages is $O(2M\langle k\rangle^4 \log \langle k\rangle \times \langle k\rangle NQ+ MQ^2)$, and if the graph is sparse the complexity becomes $O(MNQ+MQ^2)$. For relatively small $Q$, the complexity scales with $Q$ as in NDP routing as shown in Fig.~\ref{fig:EDrrComplex}.

However, when $M/N \not\ll N$, the effective degree becomes quite different from the average degree $\langle k \rangle \Rightarrow \langle k \rangle+ \frac{2M}{N}$ (see in Fig.~\ref{fig:ExtraNodeDemo}, the degree of node $i$ increases from $3$ to $4$ after introducing the auxiliary node $\mu$). When $M$ is very large, for instance $M\sim O(N^2)$ as is used in many optical communication network applications, the effect of effective degree cannot be omitted. To make the algorithm scale better we divide the messages into three types: 
\begin{enumerate}
\item messages from auxiliary nodes to ordinary nodes, e.g. $\mu\to i$ in Fig.~\ref{fig:ExtraNodeDemo} --- there are $2MQ$ such messages;
\item messages from ordinary nodes to ordinary nodes, e.g. $i\to j$ in Fig.~\ref{fig:ExtraNodeDemo} --- there are $\langle k\rangle NQ$ messages of this type;
\item messages from ordinary nodes to auxiliary nodes, e.g. $i\to \mu$ in Fig.~\ref{fig:ExtraNodeDemo} --- there are also $2MQ$ such entities.
\end{enumerate}
The messages of type 1 obeys Eq.~\eqref{eq:ND_endpoint} and the resulting total complexity is $O(MQ^2)$. 

Noticing that the messages from auxiliary nodes are sparse, only $\phi_{\mu\to a}(0)$ and $\phi_{\mu\to a}(\mu)$ are non-trivial in Eq.~\eqref{eq:ND_endpoint}, and only $\phi_{b\to \mu}(-\mu)$ is needed to generate new messages, which simplifies the calculations of Eq.~\eqref{eq:ND_mess_shortest}. Under NDP routing, by separating messages of type $1$ and $2$, the computational complexity of generating one message of type $2$ or $3$ by Eq.~\eqref{eq:ND_mess_shortest} is $O(\langle k\rangle^2 M)$. Then, the total complexity of the three types of messages is $O(MQ^2+ \langle k\rangle ^2 M(\langle k\rangle NQ+ MQ))= O(MQ(M+N+Q))$. Considering all possible pairs $M=N(N-1)/2$ and set $Q\sim O(N)$, the overall complexity becomes $O(N^5)$. In Fig.~\ref{fig:NComplex}, we present the one-iteration computing time on random regular graphs with degree $k=3$ and $N$ nodes, the one-iteration computing time scales approximately as $N^{5.14}$, which fits our analysis of $N^5$. If $M/N\ll N$, the complexity reduces to $O(MQ(N+Q))$, which is equivalent to the previous estimate of  $O(MNQ+MQ^2)$.

Under EDP routing, the computational complexity of type $2$ and $3$ messages depends on the complexity of the maximum weighted matching algorithm (approximately $O((\langle k \rangle + \frac{2M}{N})^\alpha)$) with $\alpha$ being a coefficient to be determined. Using the sparse properties of type $1$ and $3$ messages, the total complexity of the three types is $O(MQ(N+Q+(M/N)^\alpha))$. Recalling the mapping to matching problems --- the matching network has approximately $\langle k\rangle+ \frac{2M}{N}$ nodes and $\langle k\rangle(\frac{\langle k\rangle-1}{2}+\frac{2M}{N})$ edges, and the average node degree is approximately $\langle k\rangle$ when $\frac{2M}{N}\gg \langle k\rangle$, therefore it is a sparse network. By using the algorithm in~\cite{altarelli2015edge}, the complexity should be approximately $O( (\langle k\rangle+ \frac{2M}{N})^2))$. However, due to resulting topology and the special network structure our tests result in complexity scaling close to $O(\langle k\rangle +\frac{2M}{N})$ when tested. In other words, $\alpha=1$ and therefore the resulting total complexity is $O(MQ(N+Q+M/N))$. Experiments on random regular graphs presented in Fig.~\ref{fig:NComplex}, when $M=N(N-1)/2$ and $Q=N$ show that the complexity is approximately $O(N^{3.92})$ which agrees well with the analysis of $O(N^4)\sim O(MNQ)$.

\subsection{Non-linear cost}

In this section, we show the simulation results where a more general form of the cost function {$F_{i,j}(x_{i,j})= x_{i,j}^\gamma$}, where the argument is the wavelength occupancy $q_{i,j}$ on edge $(i,j)$, given by
\begin{equation}\label{eq:edge-occupancy}
x_{i,j} = \sum_{a=1}^Q (1- \delta_{s_{i,j}^a}^0) ~.
\end{equation}
When $\gamma> 1$, the utility increases faster on heavily loaded edges, biasing solutions towards paths with uniform load, balancing the loads on edges. On the other hand, when $0<\gamma< 1$, the cost on edges increases slower with the load, leading to configurations which consolidate transmissions on used edges, leaving more edges and wavelength channels idle. Such configurations would be relevant for identifying less important transceiver nodes which can be switched off in hours of low usage.

\begin{figure}[hbtp!]
\centering
\subfigure[]{
\includegraphics[width=0.226\textwidth]{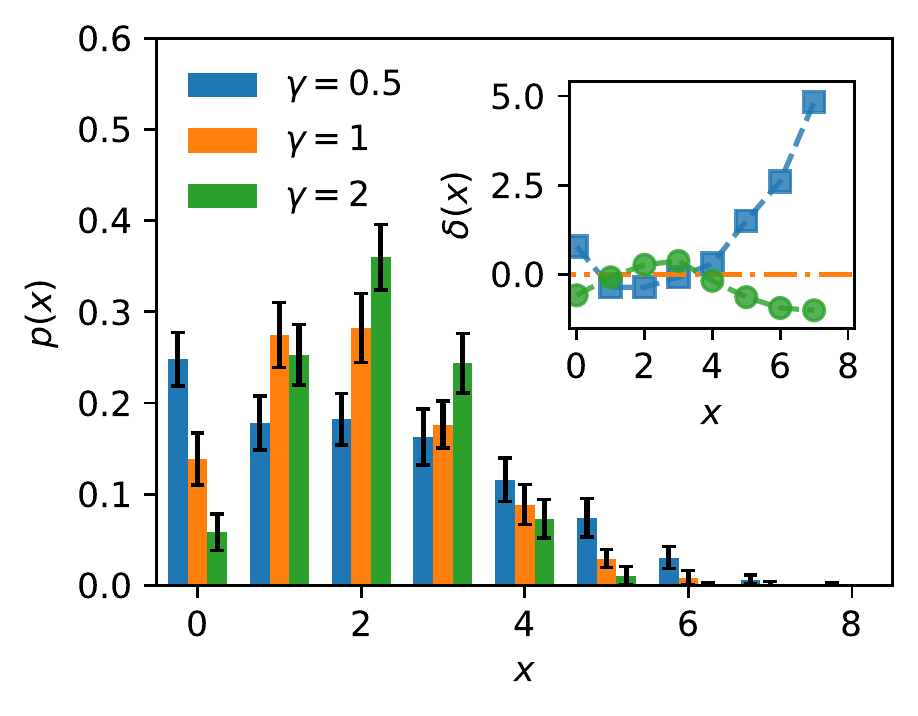}
\label{fig:gammaEffect_RR_ND}}
\subfigure[]{
\includegraphics[width=0.226\textwidth]{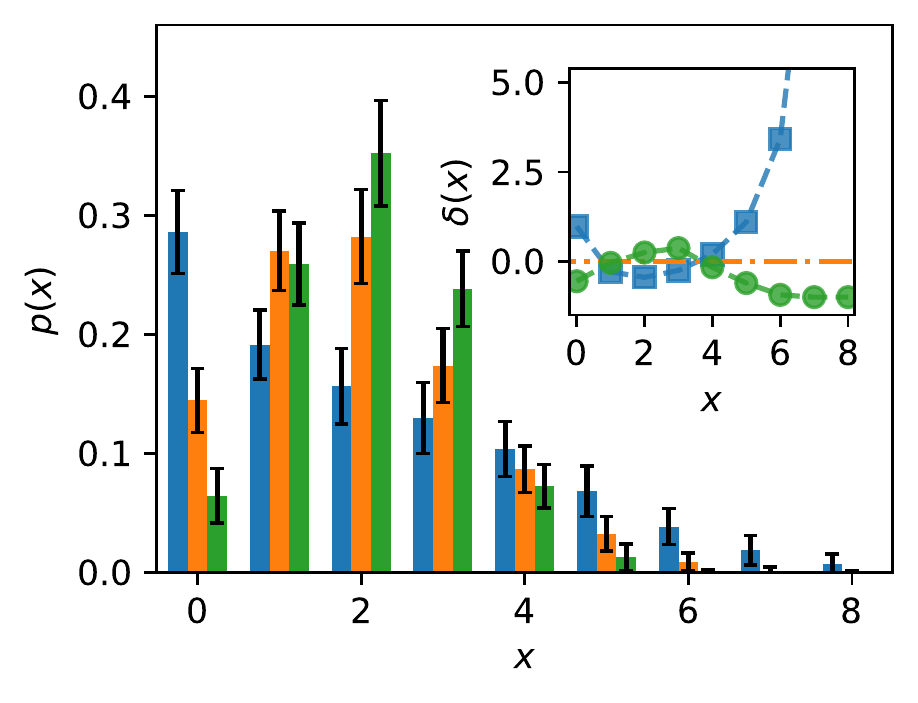}
\label{fig:gammaEffect_RR_ED}}
\subfigure[]{
\includegraphics[width=0.226\textwidth]{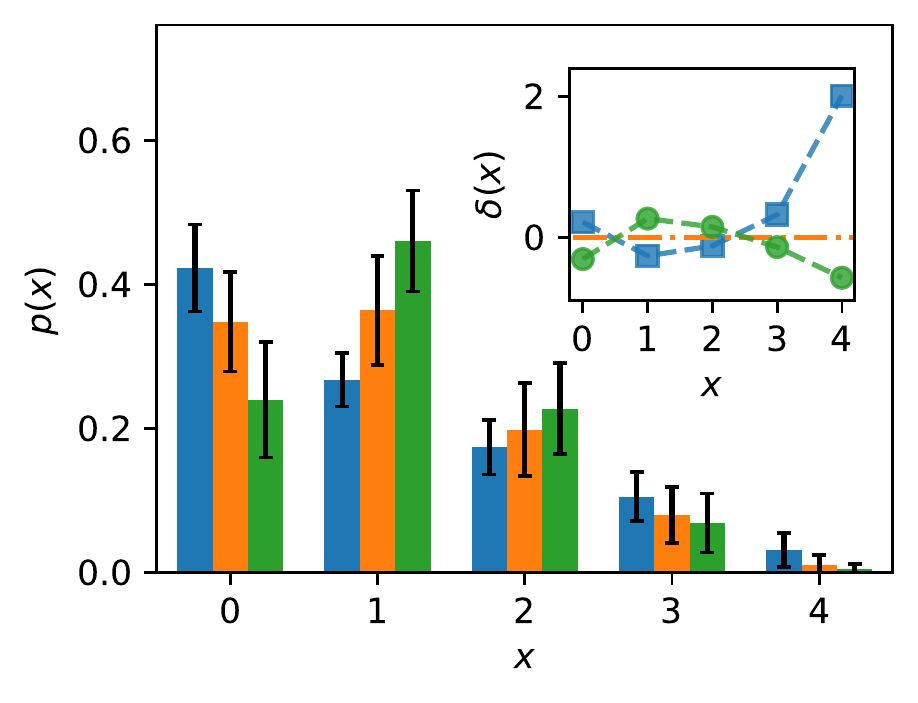}
\label{fig:gammaEffect_US_ND}}
\subfigure[]{
\includegraphics[width=0.226\textwidth]{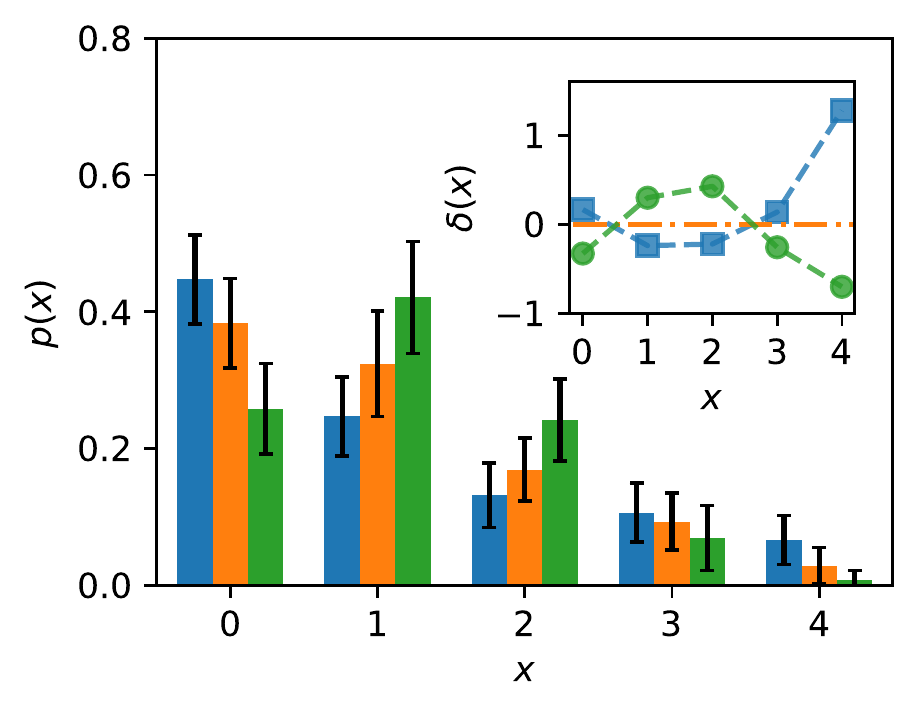}
\label{fig:gammaEffect_US_ED}}
\subfigure[]{
\includegraphics[width=0.226\textwidth]{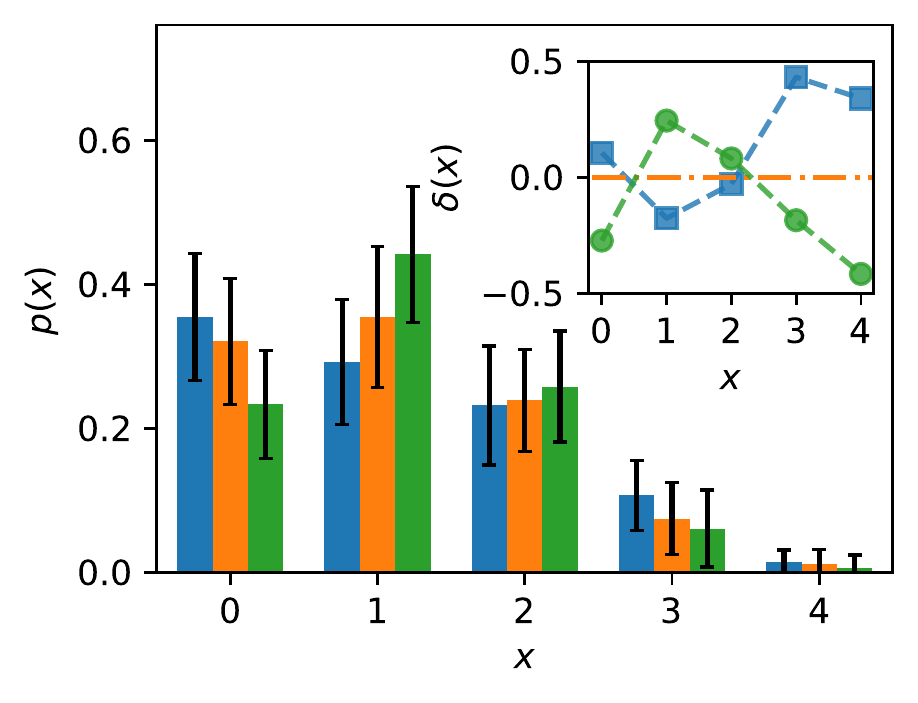}
\label{fig:gammaEffect_BT_ND}}
\subfigure[]{
\includegraphics[width=0.226\textwidth]{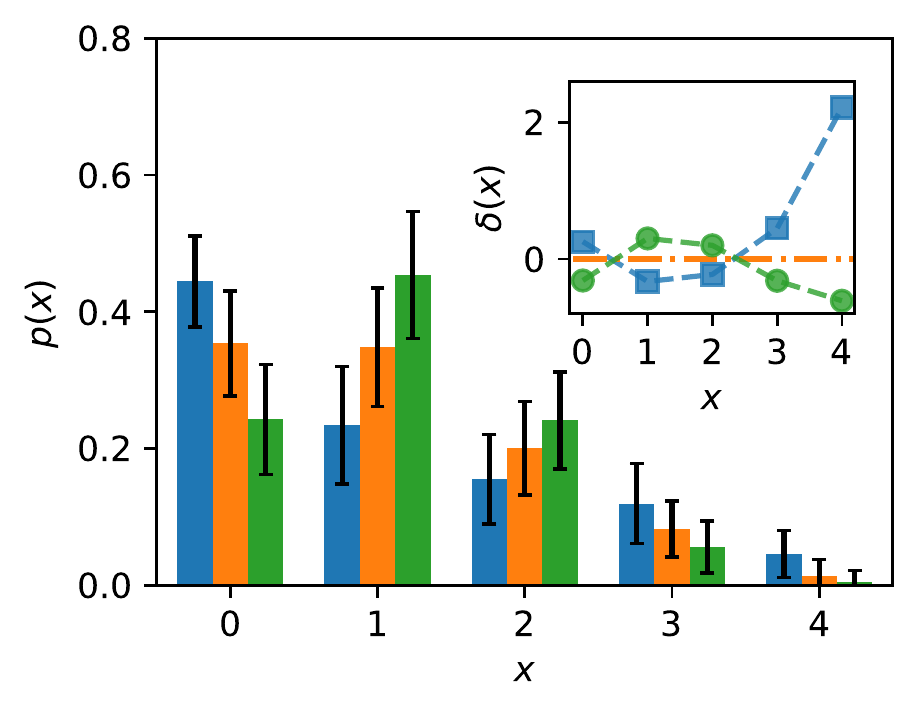}
\label{fig:gammaEffect_BT_ED}}
\caption{\label{fig:gammaEffect}The distribution $p(x)$ of wavelength occupancy on edges for power $\gamma= 0.5, 1, 2$ in the objective function; (a), (c), (e) represent NDP scenarios and (b), (d), (f) EDP scenarios for different graphs: (a), (b) random regular networks with $100$ nodes, degree $3$, $M= 60$ and $Q=8$; (c), (d) the real network CONUS with $M= 14$ and $Q=4$, and (e), (f) the BT-Core network with $M=12$ and $Q=4$. Insets: the relative difference $\delta(x;\gamma)$ between $\gamma= 0.5$ and $1$, as well as between $\gamma= 2$ and $1$. The results are obtained by averaging $22\sim 36$ realizations.}
\end{figure}

In Figure ~\ref{fig:gammaEffect} we show the distribution $p(x)$ of wavelength occupancy on edges of random regular graphs and the two real network CONUS and BT-Core for different values of $\gamma$. For all three networks and both NDP and EDP scenarios, we observe that when $\gamma= 0.5$ more edges were unused, i.e. a higher value at $p(x=0)$ in Fig.~\ref{fig:gammaEffect}, and as $\gamma$ increases the distribution become more evenly distributed and peaked at some values of $x$, which corresponds to the balancing of edge loads. We show in the insets the relative difference between $p(x)$ obtained by $\gamma=1$ and those obtained by $\gamma= 0.5$ or $2$, i.e. $\delta(x;\gamma)\equiv \frac{p(x;\gamma)-p(x;\gamma=1)}{p(x;\gamma=1)}$, where the evidence for load consolidation or balancing, for $\gamma=0.5$ or $2$ respectively, become obvious.

\subsection{Comparison with linear programming}

Linear programming is a commonly used method for optimized throughput of optical networks~\cite{ives2015routing}. Here, we compare the results of our proposed algorithm with those obtained by linear programming with a linear cost function $\gamma=1$.

To solve the routing problem using linear programming, we find $n$ shortest paths for each of the $M$ transmissions, denoted by the variables $\{\vec s_k^\mu\}$ with superscript $\mu= 1,\dots,M$ representing transmissions and subscript $k=1,\dots,n$ representing the $k^\mathrm{th}$ candidate path for transmission $\mu$. By carrying either a subscript for nodes or edges, the variables $\vec s_k^\mu$ can represent a configuration of node states or edge states, and the NDP or EDP constraints can both be expressed in terms of $\vec s_k^\mu$. In this case, we introduce a variable $\sigma_k^{a,\mu}=1$ to denote that transmission $\mu$ chooses its $k^{\mathrm th}$ path with wavelength $a$, $\sigma_k^{a,\mu}=0$ otherwise. The disjoint constraint can be expressed as follows,
\begin{equation}
\label{eq:LP_disjoint}
\forall a,j:~ v^a_j \leq 1, \mbox{ where }\vec v^a= \sum_{k,\mu} \sigma_k^{a,\mu} \vec s_k^\mu.
\end{equation} 
In Eq.~\eqref{eq:LP_disjoint}, if the path $\vec s_k^\mu$ represents a configuration of node states, then each element of $\vec v^a$ represents the load of $a^\mathrm{th}$ wavelength channel on that node, and node-disjoint constraint restricts the value of the load to be no more than $1$; if the path $\vec s_k^\mu$ represents a configuration of edges, then the edge-disjoint constraints can be defined in a similar manner.

For each transmission, we choose one wavelength to accommodate one candidate path, which is given by
\begin{equation}
\label{eq:LP_allocation}
\sum_{a,k}\sigma_k^{a,\mu}= 1,~\forall \mu=1,\dots,M.
\end{equation}
The objective function for the problem with linear cost is given by
\begin{equation}
\label{eq:LP_obj}
\mbox{Minimize } \sum_{a,j} v_j^a. 
\end{equation}
The problem can be solved by linear programming with the objective function of Eq.~\eqref{eq:LP_obj} subject to the constraints in Eq.~\eqref{eq:LP_disjoint}--\eqref{eq:LP_allocation} where all the expressions are linear.

For the WS scenario, the variables $\sigma_k^\mu$ are introduced instead of the previous $\sigma_k^{a,\mu}$, and the capacity constraint for each node is given by
\begin{equation}
\forall j:~ v_j\leq Q, \mbox{ where } \vec v= \sum_\mu \sigma_k^\mu \vec s_k^\mu,
\end{equation}
where $\vec s_k^\mu$ is a configuration of node states. The constraint for all the transmissions is given by
\begin{equation}
\sum_{k}\sigma_k^\mu= 1, ~\forall \mu=1,\dots,M.
\end{equation}
The objective function to be minimized is $\sum_j v_j$.

\begin{table}[h]
\centering
\caption{\label{tab:results_MP_LP}The smallest number of wavelength channels $Q_{\min}$ required to transmit all $M=\vert V\vert (\vert V\vert -1)/2 (=N(N-1)/2)$ transmissions on four small real optical communication networks including NSF-Net, Google-B4, DTAG/T-systems and BT-Core, obtained by our multi-wavelength routing algorithm (MP) compared to that obtained by linear programming (LP), in edge-disjoint (ED), node-disjoint (ND) and node-disjoint wavelength-switching (WS) scenarios. The corresponding total path lengths are also shown.}
\begin{tabular}{cccccc}
\hline \hline
\multicolumn{2}{c}{Network} & NSF-Net & Google-B4 & DTAG/T-systems & BT-Core \\ 
\hline 
\multicolumn{2}{c}{$|V|$} & $14$ & $12$ & $14$ & $22$ \\ 
\multicolumn{2}{c}{$|E|$} & $21$ & $19$ & $23$ & $35$ \\ 
\multicolumn{2}{c}{$M$} & $91$ & $66$ & $91$ & $231$ \\ 
\hline 
\multirow{2}{*}{MP-ED} & $Q_{\min}$ & $13$ & $16$ & $14$ & $39$ \\ 
& $L$ & $195$ & $153$ & $218$ & $697$ \\ 
\hline 
\multirow{2}{*}{LP-ED}& $Q_{\min}$ & $13$ & $16$ & $14$ & $39$ \\ 
& $L$ & $195$ & $153$ & $218$ & $697$ \\ 
\hline 
\multirow{2}{*}{MP-ND} & $Q_{\min}$ & $25$ & $23$ & $29$ & $52$ \\ 
& $L$ & $202$ & $154$ & $221$ & $707$ \\ 
\hline 
\multirow{2}{*}{LP-ND}& $Q_{\min}$ & $25$ & $23$ & $29$ & $52$ \\ 
& $L$ & $201$ & $154$ & $221$ & $709$ \\ 
\hline \multirow{2}{*}{MP-WS} & $Q_{\min}$ & $25$ & $23$ & $29$ & $51$ \\ 
& $L$ & $201$ & $154$ & $221$ &  $715$\\ 
\hline 
\multirow{2}{*}{LP-WS}& $Q_{\min}$ & $25$ & $23$ & $29$ & $51$ \\ 
& $L$ & $201$ & $154$ & $221$ & $715$ \\ 
\hline \hline
\end{tabular} 
\end{table}

We conducted numerical experiments on four real networks~\cite{ives2015routing,wright2012} and compared the results obtained by our algorithms and linear programming in the NDP, wavelength-switching and EDP scenarios. We show the smallest numbers of wavelength channels, i.e. $Q_{\min}$, needed to complete all possible transmissions, and the corresponding total path length $L$ in Tab.~\ref{tab:results_MP_LP}. As we can see, our message-passing algorithms and linear programming yield almost identical performance in finding optimized path solutions. For relatively small network, linear programming is efficient, but it quickly becomes impractical when the size of networks increases, whereas we show in Sec.~\ref{sec:complexity} that our message-passing algorithms have more practical scaling properties.

\section{Generalization to heterogeneous edge weight and wavelength availability\label{sec:extensions}}

In real optical networks, the number of wavelength channels in different optical fibers may vary; their lengths, signal-to-noise ratios, or type of fibers used may influence the quality of communication. To make our model more general and realistic, we consider the case of optical networks with weight $w_{i,j}$ and number of wavelengths $Q_{i,j}$ defined for any individual link $(i,j)$ (or $Q_i$ defined for node $i$ in node-disjoint wavelength-switching scenarios). With simple modifications, our proposed algorithm can be generalized to accommodate $w_{i,j}$, $Q_{i,j}$ or $Q_i$.

In cases with heterogeneous $Q_{i,j}$ on edges, we denote the largest number of wavelength channels among all edges to be $Q_*$, i.e. $Q_*=\max_{(i,j)}Q_{i,j}$, then for an edge $(i,j)$, one can introduce $Q_* -Q_{i,j}$ additional wavelength channels with state $0$, such that all edges on the network would have virtually $Q_*$ wavelengths. 

As for heterogeneous weights on edges with linear cost, the objective function becomes $L= \sum_{(i,j)} w_{i,j}\sum_{a=1}^Q(1-\delta^0_{s_{i,j}^a})$ and the partition function in \eqref{eq:partition} becomes
\begin{equation}
Z(\beta)= \sum_{\vec{\vec s}} \Omega(\vec{\vec s})\prod_{(i,j)}\ee^{-\beta w_{i,j}F_{i,j}(\vec s_{i,j})}.
\end{equation}

\subsection{Heterogeneous wavelength availability on edges}

In the message-passing equations \eqref{eq:ND_mess_shortest} describing NDP scenario, the first equation $\phi_{i\to j}^a(0)$ considers the condition for wavelength $a$ of edge $(i,j)$ to be in state $0$ in the absence of node $j$, and the contribution to the objective function $L$ is $0$; this equation is the same in cases with heterogeneous edge weights. Nevertheless, weights $w_{i,j}$ should be introduced in the second equation as
\begin{equation}
\phi_{i\to j}^a(s) \sim w_{i,j}+ \min_{k\in\partial i\setminus j}\Big[\phi_{k\to i}^a(s)+\sum_{l\in\partial i\setminus j,k}\phi^a_{l\to i}(0)\Big].
\end{equation}

The same applies for the EDP scenarios and the second message-passing equation in Eq.~\eqref{eq:ED_mess_short} should be modified as follows
\begin{equation}
\begin{aligned}
\phi_{i\to j}^a(s) \sim ~ w_{i,j}+& \min_{k\in\partial i\setminus j} \Big[\phi_{k\to i}^a(s)+\\
&\min_{\mbox{matched}\atop\mbox{pairs: }\vec s_{\partial i\setminus j,k}} \sum_{l\in\partial i\setminus j,k}\phi_{l\to i}^a(s_l)\Big].
\end{aligned}
\end{equation}

The marginal messages on edges, Eq.~\eqref{eq:ND_marginal_shortest}, should be  modified as
\begin{equation}
\phi^a_{i,j}(s)= \phi_{i\to j}^a(s)+ \phi_{j\to i}^a(-s)+ w_{i,j}(\delta_{s}^0- 1).
\end{equation}
The same algorithmic procedure described in Sec.~\ref{sec:algo} can be applied in the present cases with heterogeneous edge weights.

For an even more general scenarios, our model can be modified to accommodate heterogeneous weights for different wavelength channels on the same edge, i.e. $w_{i,j}^a\neq w_{i,j}^b$ for $a\neq b$, asymmetric directed weights such as $w_{i,j}\neq w_{j,i}$ on directed graphs. In these cases, one only needs to replace $w_{i,j}$ in the above modified equations by $w_{i,j}^a$ or directed weights.

\subsection{Heterogeneous wavelength availability on nodes with wavelength-switching}

For the NDP scenarios with heterogeneous wavelength availability on nodes, we do not have to introduce additional wavelengths and keep unavailable channels in state $0$ if $Q_{i}$ are non-uniform; this is because the equation of node capacity constraint Eq.~\eqref{eq:NS_capacity_constraint} has already considered the case of different $Q_i$ values. Only the second equation of Eq.~\eqref{eq:NS_mess} needs to be modified, which reads
\begin{equation}
\begin{aligned}
\phi_{i\to j}^\mu(\pm 1) \sim ~ &w_{i,j}+ \tau_i^\mu(1)+\\
& \min_{k\in\partial i\setminus j}\Big[\phi_{k\to i}^\mu(\pm 1)+\sum_{l\in\partial i\setminus j,k} \phi_{l\to i}^\mu(0)\Big].
\end{aligned}
\end{equation}

\section{Conclusion}
\label{sec:conclusion}
Multi-wavelength NDP/EDP routing lies at the heart of the efficient running and design of optical communication networks, that act as the backbone of the Internet. One of the key questions in running optical communication networks more efficiently is in the ability to carry out these routing tasks effectively for large systems. This serves for both day-to-day running of the network as well as for the design of new networks and the modification of existing infrastructure. While principled {\em single wavelength} NDP and EDP routing algorithms based on message passing have been developed already, they could not be employed in real optical networks due to the difficulty in extending the algorithms from the single wavelength to the multi-wavelength case. This essential and pivotal aspect of routing in optical communication networks makes the existing single-wavelength methods intractable and requires computational cost that grows exponentially with the system size.

To accommodate a large number of wavelengths and transmissions in large systems, we have developed algorithmic solutions that include multi-layer graphs, where each layer represents a different wavelength, and messages are passed within layer (routing assignment) and between layers (wavelength allocation). The scalable algorithm we have devised shows very good performance in manageable time scales.

We expect the algorithm to be implemented in realistic scenarios, where specific aspects of real network routing, such as heterogeneous wavelength availability and signal-to-noise ratios will have to be added in the manner outlined in Sec.~\ref{sec:extensions}. We also expect the algorithm to be utilized for network design and see several possible extensions for both localized and global message passing-based implementation. Utilization of the algorithms developed here in ad-hoc network communication, multilayer VLSI design and multilayer networks will require further study of the specific requirements for the different applications.

\section*{Acknowledgements}
DS and YZX acknowledge support from the EPSRC Programme Grant TRANSNET (EP/R035342/1) and would like to thank Caterina De Bacco for pointing us to her code available on GitHub. The work by CHY and HFP is supported by the Research Grants Council of the Hong Kong Special Administrative Region, China (Projects No. EdUHK GRF 18304316, GRF 18301217 and GRF 18301119), the Dean's Research Fund of the Faculty of Liberal Arts and Social Sciences Project No. FLASS/DRF 04418, FLASS/ROP 04396, FLASS/DRF 04624) and the Internal Research Grant (Project No. RG67 2018-2019R R4015), The Education University of Hong Kong, Hong Kong Special Administrative Region, China. DS and YZX would like to thank Ruijie Luo,  Robin Matzner and Polina Bayvel for insightful comments on practical routing in optical communication systems.

\bibliography{Disjoint-ref}

\end{document}